\documentclass[a4paper,fleqn,usenatbib]{mnras}

\usepackage[T1]{fontenc}
\usepackage{ae,aecompl}

\usepackage[latin9]{inputenc}

\usepackage{url}

\usepackage{rotating}
\usepackage{graphicx}

\usepackage{multirow}
\usepackage{amsmath}
\usepackage{amssymb}
\usepackage{bm}
\RequirePackage{color}

\usepackage{hyperref}

\graphicspath{{FiguresUsed/}}

\usepackage{float}
\newcommand{\HI}{$\mathrm{H_I}$} 
\newcommand{\Tsys}{T_{sys}}      

\newcommand{\tableline}{\hline}

\title[Design, operation and performance of the PAON4 prototype transit interferometer]{Design, operation and performance of the PAON4 prototype transit interferometer} 

\author[R. Ansari et al.]{R. Ansari$^{1,2}$,
	J.E Campagne$^{1,2}$\thanks{Email: campagne@lal.in2p3.fr},
	D. Charlet$^{1,2}$,
	M. Moniez$^{1,2}$,
       C. Pailler$^{1,2}$,	
       O. Perdereau$^{1,2}$,
	\newauthor 
    M. Taurigna$^{1,2}$,
	J.M. Martin$^{3}$,
	F. Rigaud$^{3}$,
	P. Colom$^{4}$,
	Ph. Abbon$^5$,
	Ch. Magneville$^{5}$,
	\newauthor 
        J. Pezzani$^6$,
	C. Viou$^6$,
	S.A. Torchinsky$^7$,
	Q. Huang$^{9,1}$,
	and J. Zhang$^{8,1,9}$
	\\
$^{1}$LAL, Univ. Paris-Sud, CNRS/IN2P3, Universit\'e Paris-Saclay, 91405 Orsay, France\\
$^{2}$Universit\'e Paris-Saclay, CNRS/IN2P3, IJCLab, 91405 Orsay, France\\
$^{3}$GEPI, UMR 8111, Observatoire de Paris, 61 Av. de l'Observatoire, 75014 Paris, France\\
$^{4}$LESIA, UMR 8109, Observatoire de Paris, 5 place Jules Janssen, 92195 Meudon Cedex, France\\
$^{5}$CEA, DSM/IRFU, Centre d'Etudes de Saclay, 91191 Gif-sur-Yvette, France\\
$^{6}$Station de Radioastronomie de Nan\c{c}ay, Observatoire de Paris, PSL Research University, CNRS, Universit\'e d'Orl\'eans,\\ 18330 Nan\c{c}ay, France\\
$^{7}$APC, Universit\'e  Paris Diderot, CNRS/IN2P3, CEA/Irfu, Observatoire de Paris, Sorbonne Paris Cit\'e, 75205 Paris Cedex 13, France\\
$^{8}$College of Physics and Electronic Engineering, Shanxi University, Shanxi 030006, China\\
$^{9}$National Astronomical Observatories, Chinese Academy of Sciences, Beijing 100012, China 
}

\date{Accepted 2020 February 03; Received 2020 February 03; in original form 2019 October 11}

\pubyear{2020}

\begin{document}
\label{firstpage}
\pagerange{\pageref{firstpage}--\pageref{lastpage}}
\maketitle
 
\begin{abstract}
PAON4 is an L-band (1250-1500 MHz) small interferometer operating
in transit mode deployed at the Nan\c{c}ay observatory in France,
designed as a prototype instrument for Intensity Mapping.  It features
four 5~meter diameter dishes in a compact triangular configuration, 
with a total geometric collecting area of $\sim75 \mathrm{m^2}$, and 
equipped with dual polarization receivers. 
A total of 36 visibilities are computed from the 8 independent RF signals by the 
software correlator over the full 250~MHz RF band. 
The array operates in transit mode, with the dishes pointed toward a fixed declination, 
while the sky drifts across the instrument.  Sky maps for each frequency channel are then 
reconstructed by combining the time-dependent visibilities from the different
baselines observed at different declinations.  
This paper presents an overview of the PAON4 instrument design and goals, as a prototype for dish arrays to map the Large
Scale Structure in radio, using intensity mapping of the atomic hydrogen $21~\mathrm{cm}$ line.  
We operated PAON4 over several years and use data from observations in different periods 
to assess the array performance.  We present preliminary analysis of a large fraction of this data and discuss 
crucial issues for this type of instrument, 
such as the calibration strategy, instrument response stability, and noise behaviour. 

\end{abstract}




 
\section{Introduction}
\label{Sec:Intro}
Despite the tremendous success of the $\Lambda$CDM cosmological model
and the wealth of information provided by the analysis of CMB
anisotropies \citep{2013ApJS..208...19H, 2016A&A...594A..13P,
  2018arXiv180706209P}, SNIa luminosities \citep{Betoule:2014frx} and optical surveys \citep[see
  e.g.][]{2017MNRAS.468.2938S, 2018MNRAS.tmp.3191A,
  2018arXiv181102375D}, the nature of Dark Matter and Dark Energy
remain elusive \citep{2016PDU....12...56B} and a number of fundamental
physics questions still need clarification.  The Large Scale Structure
(LSS) and its Baryon Acoustic Oscillation (BAO) feature are among the
most powerful probes to constrain the cosmological model and Dark
Energy and Modified gravity models \citep{2013PhRvD..87b3501A}.  The
matter distribution at large scale in the universe is usually mapped
through optical photometric and spectroscopic surveys.

The 21~cm line emission/absorption at 1.42~GHz due to the hyperfine
transition of neutral atomic hydrogen \mbox{(\HI)} is a unique spectral
feature in the L/UHF band which can be used to trace matter distribution 
from the vicinity of the Milky~Way at very low redshifts, up to cosmological
distances at $z \sim 1-3$, and into the Epoch of Reionization (EoR) at redshifts well
above~10.  Using the 21~cm line to study the dark ages and the EoR has
been considered for more than 20~years \citep[see
  e.g.][]{2006PhR...433..181F,2008PhRvD..78j3511P} and several
experiments have been carried out or ongoing to detect the 21~cm signal from
the EoR.  {\color{black}LOFAR \citep{2013A&A...556A...2V} , MWA
\citep{2013PASA...30....7T} , PAPER \citep{2010AJ....139.1468P} , HERA
\citep{2014ApJ...782...66P}, NenUFAR \citep{zarka:hal-01196457} are among the main dedicated 
instruments targeting EoR signal detection}.

As noted already more than a decade ago
\citep{2005MNRAS.360...27A,2006astro.ph..6104P}, mapping the
3D~distribution of matter using \HI\ as a tracer is an elegant and
complementary approach to optical surveys to constrain cosmological
parameters and Dark Energy through the study of LSS and BAO.  The
early proposals considered the observation of individual
galaxies at  $\lambda=$21~cm, requiring very large collecting area and
sensitivities, which is expected for SKA\footnote{SKA organisation: \tt{https://www.skatelescope.org/}} 
\citep{2015aska.confE.174B}.

However, it was soon realized that the study of LSS with the 21~cm emission could be
carried out through the Intensity Mapping technique, without the need
to resolve individual \HI\ sources, and with less stringent
requirements on instrument sensitivity.  To perform such a survey, a
wide band instrument with a wide field of view is required, operating in the
L/UHF band, from a few hundred~MHz to 1400~MHz $(\nu = 1420.4/(1+z) \,
\mathrm{MHz})$.  A modest angular resolution of a fraction of a
degree ($\sim 10'$) is enough to determine the matter power
spectrum on the scales most useful for Cosmology
\citep{2008PhRvL.100i1303C,2008arXiv0807.3614A,2008MNRAS.383.1195W,2009astro2010S.234P}.

This approach has a major drawback 
even without considering the instrument noise and Radio Frequency
Interference (RFI).  
The foreground emission in the radio spectrum, around 1~GHz, 
has brightness temperatures above $10\mathrm{K}$, and is dominated by the Milky Way 
synchrotron emission and radio sources. The foregrounds are thus a few
thousand times brighter than the cosmological signal which is of the
order, or below, $1~\mathrm{mK}$. This huge level of foreground emission 
makes the component separation a challenging task, much more difficult  compared to CMB observations. 
The smooth frequency dependence of
the synchrotron emission is the key to solving this daunting task. The
problem is similar for experiments looking for the 21~cm signal from
the EoR and has been studied by a number of authors
\citep[e.g.][]{2006ApJ...650..529W}.

The detectability of the cosmological 21~cm signal at lower redshifts
($z \lesssim 3$) in the presence of foregrounds has been subsequently
studied \citep[e.g.][]{A&A...540A.129A,2014MNRAS.441.3271W}. The
advantages of the m-mode analysis for map making and foreground
subtraction, applicable for a transit type interferometer is discussed
in \cite{2014ApJ...781...57S} and extended to the polarization case
in \cite{2015PhRvD..91h3514S}.
  
The interest in Intensity Mapping (IM) as a tool for cosmological and
Dark Energy surveys has grown over the last decade, with broadened
science reach \citep{2015ApJ...803...21B,2018arXiv181009572C}.
A~number of groups have engaged in exploiting existing instruments to
carry Intensity Mapping surveys, for example using GBT
\citep{2010Natur.466..463C,2013ApJ...763L..20M} or plan to do
so with SKA and its pathfinders \citep{2015aska.confE..24B}, or with
single-dish telescopes \citep{2015MNRAS.454.3240B} such as FAST 
\citep{2016ASPC..502...41B}. 

Other groups are building specifically designed instruments for such
surveys. Most groups plan to use dense interferometric arrays,
associated with high throughput digital beam-forming and digital
correlators that can fulfil the instrumental requirements for IM
surveys.  Single dish instruments combined with multi-feed receivers
provide an alternative instrumental design. A 7-beam receiver covering
the frequency band $700-900 \, \mathrm{MHz}$ was envisaged for the GBT  
\citep{2016AAS...22742601C} and a $\sim50$ feed receiver will be built
and placed at the focal plane of an off-axis reflector by the BINGO
team \citep{2013MNRAS.434.1239B, 2018arXiv180301644W}.

Dedicated IM instruments share a number of common features and problems with wide field arrays built or under development 
to search for the EoR signal, {\color{black} such as LOFAR, MWA, PAPER, HERA, and NenUFAR, although the latter observe at much lower frequencies $(\nu \sim 100~\mathrm{MHz}, z \sim 10)$}. Two main options are considered for IM interferometric arrays. Either cylindrical reflectors, oriented north-south, possibly with FFT beam-forming along the cylinder axis, or an array of dishes, with pointing capability in declination. 
FFT beam-forming is also possible using a regular grid of antenna, including dishes, as mentioned by \citep{2009PhRvD..79h3530T}.  Such a strategy has already been implemented on a modest scale in individual LOFAR stations for the High Band Array, and at around 1~GHz frequencies in EMBRACE \citep{2016A&A...589A..77T}.  This was the ultimate goal of the SKA mid frequency array with an early demonstrator called 2-PAD \citep{2009wska.confE..45A}. 

{\color{black}In addition to the cosmological 21~cm signal detectors, such dedicated instruments will be powerful machines 
for time domain radio astronomy. Indeed the large mapping speed of transit interferometers would make them suitable to search 
for rare transient phenomena, such as Fast Radio Bursts (FRB) or pulsars. FRB's correspond to bright broadband 
radio flashes\footnote{The FRB catalog: {\tt http://frbcat.org} }, 
discovered in recent years \citep{2013Sci...341...53T}. High values of dispersion measures (DM) suggest an extragalactic origin,
confirmed by the host identification for at least one FRB \citep{2017Natur.541...58C}. 
Pulsars can be considered as astrophysical clocks, and high precision timing measurements of 
a large number of millisecond pulsars (MSP) is being carried out in Pulsar Timing Array (PTA) programs \citep{2019MNRAS.490.4666P},  
to search for low frequency gravitational waves. Dedicated searches are also being carried out, such as \citep{2014ApJ...791...67S}
to increase the MSP population. Wide band transit interferometers could also carry FRB and pulsar searches in parallel to the
intensity mapping survey,  if equipped with dedicated digital backends.}

CHIME \citep{2014SPIE.9145E..22B}, which is among the most advanced IM projects, uses an array of cylinders, as pioneered by the 
Pittsburgh Cylindrical Telescope (PCT) prototype  \citep{2011PhDT.......158B}. The HIRAX project \citep{2016SPIE.9906E..5XN} 
plans to build a large array of 6~m dishes, in South Africa, close to the SKA site, and larger arrays, such as the proposed PUMA 
are being considered \citep{2019arXiv190712559B}.

Development of the PAON4 array and the data analysis procedures is
closely linked with Tianlai.  Tianlai is an
NAOC-led project \citep{2018SPIE10708E..36D,2012IJMPS..12..256C} which
is currently operating two pathfinder instruments in a radio quiet
site in the Xinjiang province of western China. The first Tianlai
instrument is composed of three cylinders, each $40\mathrm{m}$ long
and $15\mathrm{m}$ wide, equipped with a total of 96~dual
polarization receivers.  Following initial evaluation and
design work for PAON4, the Tianlai instrument concept was extended by
including a second array (Tianlai16D) composed of
16~steerable, 6~meter diameter dishes, in a dense circular
configuration. We carried out studies to optimize the PAON4 and Tianlai16D 
array configurations and evaluate performance
\citep{2016RAA....16..158Z,2016MNRAS.461.1950Z}.

This paper is organized as follows.  The PAON4 instrument design and
project history is presented in Section~2.  Section~3 presents an
overview of the electronics chain, the software correlator and the
acquisition system.  The instrument operation and the survey are
described in Section~4, while the data analysis pipeline is discussed
in Section~5.  Section~6 presents the instrument performance and some
preliminary results from the ongoing survey.  The paper concludes with
an outline of our future plans for PAON4 followed by concluding
remarks.

  
\begin{figure*}
\centering
\includegraphics[width=0.9\textwidth]{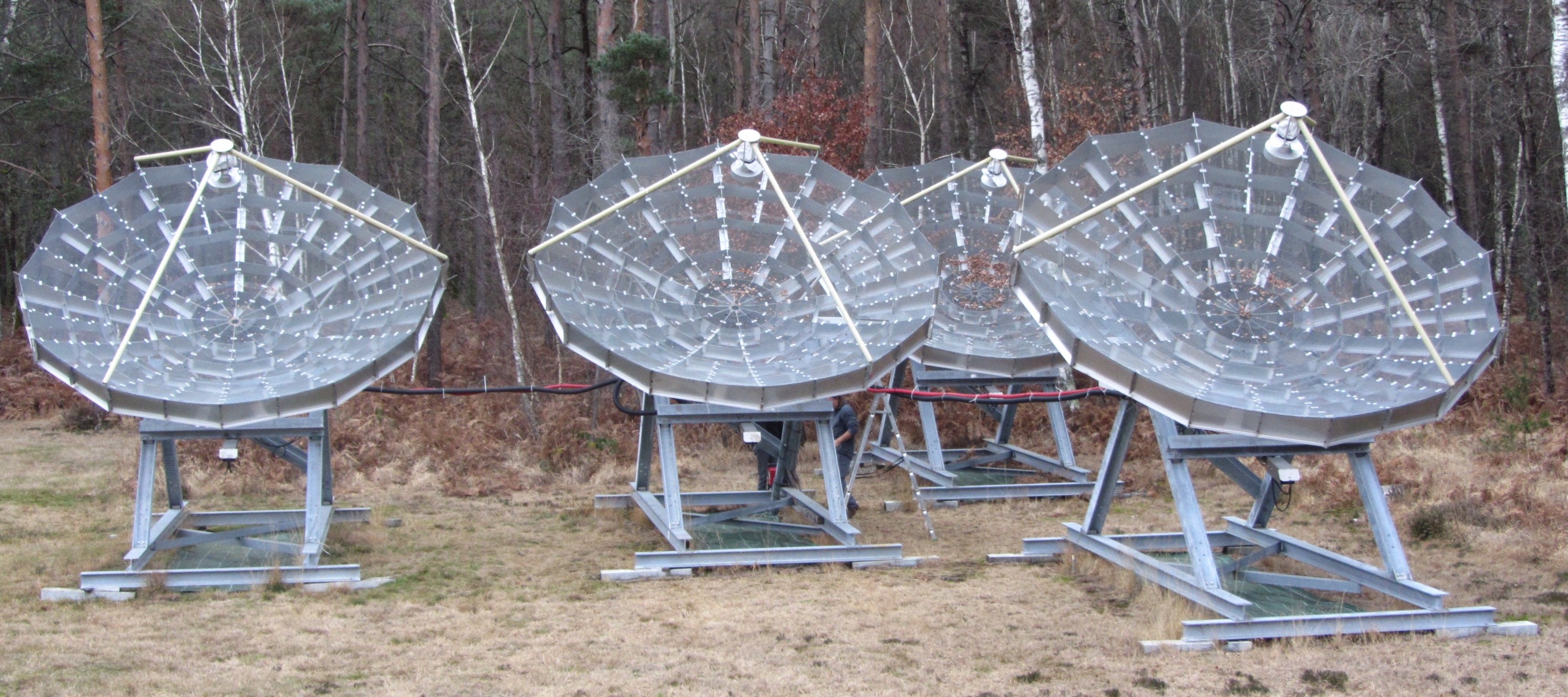} \\[1mm]
\caption{PAON4 dish array interferometer at  Nan\c{c}ay, viewed from the south-west. }
\label{Fig:paon4gview}
\end{figure*}
\section{PAON4 instrument description}
\label{Sec:paon4instrument}
Preliminary design of a small dish interferometer at Nan\c{c}ay began
in 2010.  The project, subsequently named \textbf{PA}raboles \`a
l'\textbf{O}bservatoire de \textbf{N}an\c{c}ay (PAON - Paraboloids at
the Nan\c{c}ay Observatory), was approved in spring~2012 by the partner
institutes:~~LAL-CNRS/IN2P3, Observatoire de Paris, and Irfu/SPP
(CEA). The goal is to evaluate specific issues of dense
interferometric arrays operating in transit mode.  In particular, the
project investigates:
\begin{itemize}
\item Electromagnetic coupling between neighbouring receivers at close distance and the impact on the correlated noise 
\item Array stability and the gain and phase calibration using bright sky sources 
\item Data analysis and sky map reconstruction from time dependent visibilities 
\end{itemize}

During a short test period from autumn~2012 to summer~2013 two
3~m diameter dishes were used for observation at Nan\c{c}ay.  Meanwhile,
the PAON4 reflectors were designed and manufactured. The mount
and reflectors were assembled on site during the end of spring/summer
2014 and then equipped with the receivers and electronic modules in
autumn~2014.  First observations were carried out in winter~2015, and
the instrument was formally inaugurated in April~2015.  The four
reflectors are visible in the photograph shown in
Figure~\ref{Fig:paon4gview}, taken from a point south-west of the array.
\subsection{Array configuration}
The PAON4 interferometer is composed of four 5~m diameter parabolic
$\mathrm{F/D}=0.4$ reflectors and has a 
combined {\color{black} geometric collecting area of $\sim78\mathrm{m^2}$
  ($\sim60\mathrm{m^2}$ effective area for $D_\mathrm{eff}\sim4.4\mathrm{m}$)}. 
It is hosted at the Nan\c{c}ay radio observatory\footnote{Nan\c{c}ay radio observatory, {\tt https://www.obs-nancay.fr/}}, 
in France, about 200~km south of Paris. The array (central antenna) is located at latitude 
$47^\circ 22' 55.1" \, \mathrm{N}$, and $2^\circ 11' 58.7" \, \mathrm{E}$ longitude.  

\begin{figure}
\centering
\includegraphics[width=0.6\columnwidth]{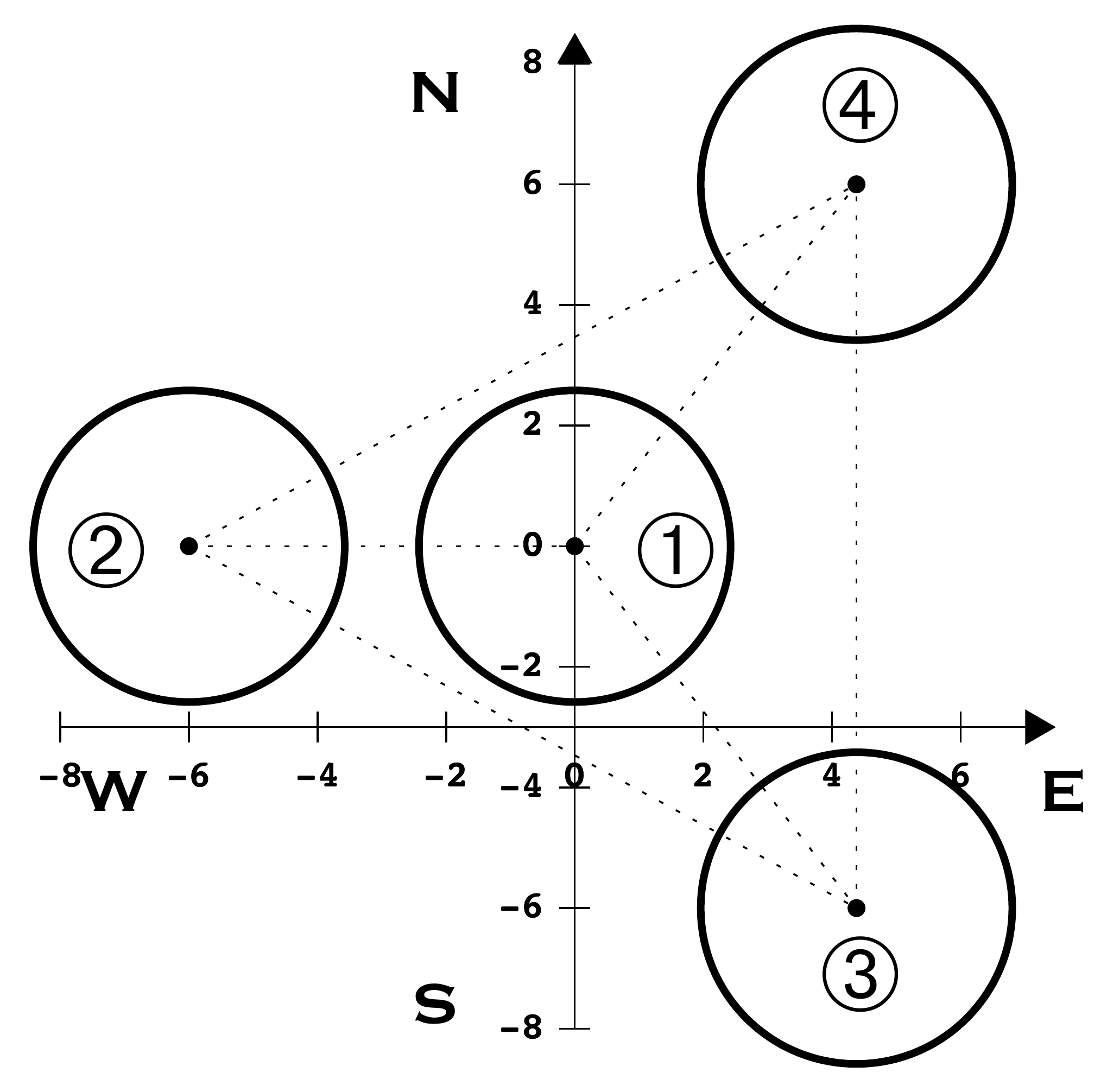} \\[1mm]
\caption{PAON4 array layout (coordinates are in meters). }
\label{Fig:paon4layout}
\end{figure}

The PAON4 compact triangular configuration was selected after
comparison of several possibilities.  The chosen configuration
provides 6 independent baselines, which is the maximum possible with 4
antennas, compared to 4 baselines for a rectangular $2 \times 2$
configuration. {\color{black} Moreover, the beam of the triangular
  configuration has a much more circular shape than the $2\times2$,
  and its resolution is also slightly better than a single dish of
  14~m diameter} \citep{2016MNRAS.461.1950Z}.  Three dishes are
positioned at the three vertices of an equilateral triangle with 12~m
sides, while the fourth dish is located close to the center of
the triangle. In addition to the four auto-correlations signals (zero
length baseline), there are six independent baselines, as shown in the
array layout schematic of Figure~\ref{Fig:paon4layout}.  One baseline
(1-2) is aligned with the East-West direction, one baseline (3-4) lies
along the North-South direction, and the four other baselines have
both EW and NS components.
 
The receiver on each reflector is made of a cylindrical feed with two
probes sensitive to two polarizations.  The H-probe is sensitive to
the electric field parallel to the EW direction, and the V-probe is
sensitive to the field in the meridian plane. There are thus a total
of 8~RF signals from the four feeds:~~Four H-polarization signals and
four V-polarization.  These are amplified, filtered, and frequency
shifted to VHF band before being digitized. The digitized signals are
processed in real-time by a a small cluster of acquisition computers
which calculate all correlations.  Two additional RF signals 
pass through identical analogue and digital electronic chains before being fed 
to the acquisition computer.  The first one,  called RFIMON, carries an RF signal from a
a simple dipole antenna and is used to monitor the RFI environment in the PAON4 band. 
The second RF signal comes from a 50~$\Omega$ resistor connected to the input of a LNA, identical to
the one which equips the H and V probes. This RF signal, referred to
as THERMON, is used to monitor the temperature dependent gain
variations of the analogue electronic chain and correct for it.  This
is discussed further in Section~\ref{Sec:Data-processing}.  PAON4 also includes  a
standard temperature probe which is read every few minutes, providing
temperature variations of the antenna environment.
The mount, reflector and feed design is discussed hereafter, and the
electronic chain and the software correlator are described in Section~\ref{Sec:electronicandcorrelator}


\subsection{Reflectors and feed design}

The antennas were designed and built at the Observatoire de Paris GEPI laboratory and by industry subcontractors. 
The design has been chosen in order to fulfil the following requirements: 
\begin{itemize}
\item build four prototype antennas with 5~m diameter with a parabolic reflector;
\item build a robust antenna architecture to be able to sustain
  historical storm conditions in the region of Nan\c{c}ay (i.e. wind speed
  of $\sim130$~km/h) as well as snow and ice during severe winter
  conditions;
\item permit the dish motor to orient in a broad range of elevation (declination) angles;
\item limit as much as possible the cost while ensuring reliable operations for 10~years or more.  
\end{itemize}
Some design parameters are summarized in
Table~\ref{Tab:Paon4-meca-caract} and the whole array is shown in
Figure~\ref{Fig:paon4gview}.

The mechanical structures of the antennas are standard steel structure
elements which underwent anti-corrosion treatment by hot-dip
galvanizing, and were prepared (cut, drilled) by numerically
controlled machines by the subcontractor. Each parabola is made of
12~petals for logistic reasons.  Each petal is made of a laser-cut
stainless steel sheet with a carefully hand-installed reflector
mesh. About 600~hours were required to install the mesh at the GEPI
laboratory. The surface deviations of the mesh were kept as low as
possible, and they were measured after mounting at a level of
$\lambda$/70~r.m.s..  This design is well adapted for a project on the
scale of PAON4, but for a larger array a more automated process would
be required since non-negligible manpower on site was required for the
mounting of the 48~petals.

Asynchronous motors with reduction gears and non-reversible
trapezoidal screws are used for the declination axis
pointing. The rotation of the motor is selected by embedded
software, and takes into account the stop phase duration for better
pointing precision.  The position of the parabola is controlled by an
accelerometer which is installed in a sealed control command
box and is therefore fully protected against the weather conditions,
unlike the position encoders. In addition to performing pointing
measurements, the accelerometer can monitor any small angular
displacement of the antennas such as those due to possible
ground instabilities.  A running mean of 2000~samples is performed in
order to measure the position of each parabola. The pointing of the
antennas is controlled by scripts or by an embedded web server accessible
through the internet. A precision of $\sim0.1^{\circ}$ has been
measured on site.

The PAON4 antennas are installed near the EMBRACE prototype
\citep{2016A&A...589A..77T}, and could therefore be easily connected
to the electric mains and Ethernet network. At this place, the ground
is particularly loose and in order to avoid having to build expensive and fixed
concrete foundations, the antenna structures are slightly flexible
with respect to horizontal stability, and are equipped with four legs
in a statically undetermined configuration. In addition, the rotation
axis bearings include kneecaps.
\begin{table}
\small
\caption{Main design parameters of PAON4 dish and feed. }
\label{Tab:Paon4-meca-caract}
\centering
  \begin{tabular}{c c}
   \tableline
   \multicolumn{2}{c}{Dish}\\
   \tableline
   Diameter reflector & 5~m \\
   f/D          & 0.4 \\
   L x W x H   & 6~m x 5~m  x 5.4~m \\
   roughness(design) & $\lambda/50$ at 21~cm\\
   zenith angle     & $15^\circ$N to $38^\circ$S\\
   Total weight & {\color{black} 2200~kg}\\
   \tableline
      \multicolumn{2}{c}{Feed \& choke \& probes}\\
         \tableline
      Diameter  ``choke''  & 360~mm \\
      Height ``choke'' & 80~mm \\
      Distance from entrance & 10~mm\\
      Thickness & 1.5~mm\\
      Diam. internal ``feed'' & 158~mm \\
      $\Delta D/D$ & $5.\ 10^{-3}$ (0.8~mm)\\
      Thickness & 4.5~mm \\
      Height (total)                 & 377~mm \\
      Probes & 2 (linear)\\
      Probe (Diam./Length)& 6.2~mm/50.2~mm\\
      Conic part of the probe & 2.5~mm\\
      Distance probe-end cap & 79.2~mm\\
      Base & N std.  Radiall R 161 404 \\   
      Gain (meas./simu)  & 9.5~dB/10.34~dB  \\
  \tableline
   \end{tabular}
\end{table}
%

\begin{figure}
\centering
\includegraphics[width=0.45\columnwidth]{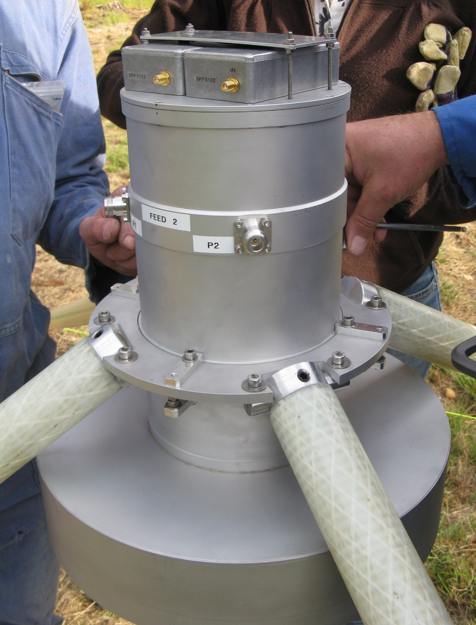}
\caption{
Photograph of a feed: The two amplifiers for H and V probes are located in the two boxes at the top.}
\label{Fig:feedpaon}
\end{figure}

The feed design was adapted to the dish diameter, taking into account
the needs of two orthogonal linear polarizations and the total
bandwidth $[1250, 1500]$~MHz.  It uses a classical circular wave guide
with a choke.  The geometric parameters are listed in
Table~\ref{Tab:Paon4-meca-caract} and shown in the sketch of
Figure~\ref{Fig:feedpaon} (left panel). The guide has a diameter of
158~mm and a length of 377~mm machined inside a aluminium guide 4.5~mm
thick. The cut-off frequencies are 1110~MHz for the TE11 mode and
1460~MHz for the TM01 mode making the feed system single moded
throughout the bandwidth except in the upper 40~MHz. The guide
wavelength is $\lambda = 543, 341, 299$~mm at $1250, 1420,
1500$~MHz. The machining guarantees a circular tolerance of $\Delta
D/D = 5\times10^{-3}$ (0.8~mm) for a cross-polarization isolation of
$-30$~dB.  This feed is suitable for the PAON4 bandwidth but could not
be easily adapted to a larger bandwidth.  The probes are made with
brass rods of relatively large diameter to have the possibility of
larger bandwidth in the future.  The conic shape of the rod minimizes
the impedance mismatch with the base. The exact locations of the
probes inside the circular guide have been optimized by simulation and
verified by measurement \textit{a posteriori}. The choke design has 
also been optimized by simulation to reduce side lobes as much as 
possible.

\begin{figure}
\centering
\includegraphics[width=0.75\columnwidth]{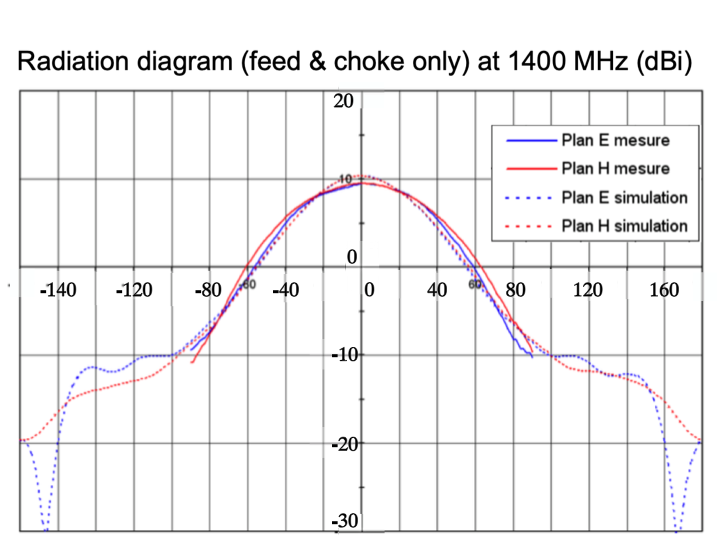}
\caption{Radiation diagram of the feed and choke: measurement (solid lines) and simulation (dashed lines). }
\label{Fig:feed_raddiag_mes}
\end{figure}

Figure~\ref{Fig:feedpaon} (right panel) shows the manufactured feed. 
The feed is maintained at the focal location by three fibreglass hollow bars 
to optimize the electromagnetic transparency of the
supports while ensuring robustness during high wind conditions. The
radiation diagram of the feed alone has been simulated and then
measured with good agreement, as shown in
Figure~\ref{Fig:feed_raddiag_mes}.


\section{Electronic and acquisition system}
\label{Sec:electronicandcorrelator}

PAON4 is equipped with the analogue and digital electronic modules
developed in the BAORadio project which began in 2007, mostly as
digital components for a 21~cm \HI\ Intensity Mapping survey
\citep{2012CRPhy..13...46A}.  The ADC boards are based on FPGA
technology.  Each is capable of digitising 4~RF signals at 500~MHz
sampling rate, and performing an FFT in real time with 61 kHz
frequency resolution \citep{2011ITNS...58.1833C}.

An analogue RF amplification, filtering, local oscillator, and mixer
were designed and realized for testing the digital components boards
with the PCT (Pittsburgh Cylindrical Reflectors).  The LNA was
specifically designed and made for PAON4.  The electronic chain, and
in particular the analogue components, were designed primarily for
testing purposes and the performance does not equal the current
state-of-the-art.

The BAORadio electronics and the associated data acquisition software
were used several times on the PCT at Pittsburgh
\citep{2011PhDT.......158B}.  In 2010, 32 signals were processed from
the PCT feeds, including the computation of the full set of 512
correlations, over a reduced band of $\sim 50 \mathrm{MHz}$.  The same
system was also used on the FAN project \citep{6588971} and
successfully deployed at the NRT (Nan\c{c}ay Radio Telescope)
for the HICluster program, searching for \HI\ in nearby clusters
\citep{2016ExA....41..117A}.


\subsection{Analogue and digital electronic}
\label{Sec:analogelectronic}

\begin{figure}
\centering
\includegraphics[width=\columnwidth]{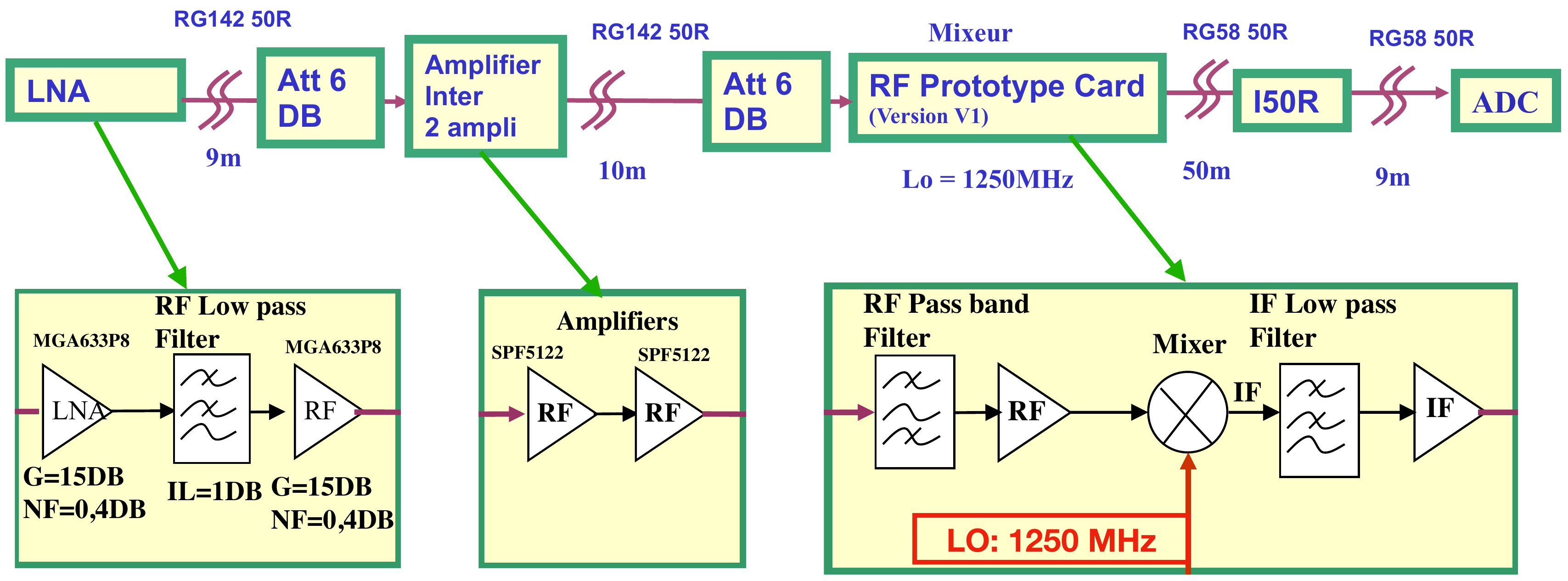} \\[1mm]
\caption{Schematic view of the PAON4 analogue electronics chain} 
\label{Fig:analogelec}
\end{figure}

An overall schematic view of the PAON4 analogue electronics chain is shown in Figure~\ref{Fig:analogelec}. 
%
%

The amplifier circuit consists of two LNA's
interconnected through a low pass filter. The amplifiers are installed at the feed.  They are connected to the
probes through $10\;\mathrm{cm}$ cables.  An Avago/Broadcom MGA633
GaAs MMIC ultra low noise amplifier is followed by a Qorvo SPF5122
LNA, and a low pass filter is inserted between the two LNAs to lower
the power at the input of the second one.  The total gain is $30\;dB$
and the noise factor $\mathrm{NF} \sim 0.6 \;dB$, leading to a noise
temperature of $T_a \sim 50 \mathrm{K}$ for the the first stage
amplifier ($F = 1 + \frac{T_a}{290 \mathrm{K}} ; \mathrm{NF} = 10
\log_{10}(F) $).
An $8\;\mathrm{m}\;RG142$ cable connects the feed amplifiers to
the second stage amplifier on the antenna.  This stage consists of two
SPF5122 LNA and provides an additional gain of $30\;dB$, which is
connected to the frequency shifter in the main analogue electronics
cabinet on the central antenna with a second
$8\;\mathrm{m}\;RG142$ cable.  After an RF filter, the signal is
amplified before down conversion.  An {\color{black} ADF4360--5} PLL and
VCO frequency synthesizer delivers a $1250\; \mathrm{MHz} $ signal to
a {\color{black} JMS--11X} frequency mixer.  The signal is down
converted to the $\left[ 0-250 \mathrm{MHz} \right]$ VHF band.
Additional VHF amplification is performed and a low pass filter
($20\;dB$ at $1500\; \mathrm{MHz} $) eliminates the high frequency
part of the signal.  The RF card output is connected to the ADC board
in the EMBRACE \citep{2016A&A...589A..77T} container by two
$50\Omega\;RG58$ cables.  A first $40\;\mathrm{m}$ cable brings
the signal to the feedthrough panel at the container wall, followed by
a $10\;\mathrm{m}$ cable from the feedthrough panel to the PAON4
digital electronics.

Each BAORadio ADC board includes two (ADC+FPGA) blocks sharing common
control electronics.  Each block handles two analogue inputs which are
sampled at 500~Msample/s with 8-bit dynamic range.  Each block
transmits data through high speed optical links to dedicated
PCIExpress reception boards in the acquisition computers.  The eight
PAON4 RF signals (4 H-pol + 4 V-pol) are thus digitised by two ADC
boards, and data is transmitted via 4 optical fibres to the
acquisition computers located a few hundred meters away. A third ADC
board handles the RFIMON and THERMON channels, which are treated
separately by the acquisition computer and do not go through the
correlator.  
The ADC boards are configured and controlled through USB links  and they receive a common master clock and trigger signals 
through dedicated connectors.  
Two different versions of the firmware can be loaded on the ADC board FPGA's. 
The first one, called the RAW firmware is used
to digitise signal and transmit waveforms to the acquisition
computers. The RAW firmware can handle a maximum of 48k samples
corresponding to $96~\mathrm{\mu s}$ long digitization frames.  The
second firmware, called the FFT firmware, performs waveform sampling,
Fourier transform (FFT), clipping, and then transmits the Fourier
coefficients (2 byte complex numbers) for 8k sample digitization
frames corresponding to $16~\mathrm{\mu s}$ in time and 4096 Fourier
coefficients with 61~kHz frequency resolution.  Each digitization
frame is time-tagged with the 125~MHz FPGA clocks and the digitization
frames on different FPGA's are synchronised by the external trigger
signal. The BAORadio ADC boards can cope with a maximum trigger rate
of $\sim 30 \, \mathrm{kHz}$, corresponding to $\sim 50 \%$ on-sky
time.  However, the current acquisition computer system and software
correlator can only cope with a trigger rate of $\lesssim 5 \,
\mathrm{kHz}$, corresponding to $\lesssim 10 \%$ on-sky time.  We
carried out observations with the FFT firmware in September and November 2016 (see Table~\ref{Tab:listeobs}),
but the results discussed in this paper are mostly based on data taken
with the RAW firmware.  The digital data stream, either waveform or
Fourier components are transferred over optical fibres to the software
correlator-acquisition computer cluster located $300\;\mathrm{meters}$
away, in the Nan\c{c}ay computer building.

\subsection{Data acquisition and software correlator}
The data acquisition system is composed of a cluster of computers
located in a dedicated room in Nan\c{c}ay.  They are interconnected
through a private 10~Gbit/s Ethernet network.  Communication to the
outside world is done through the external 1~Gbit/s Ethernet network.
An additional computer connected through the Nan\c{c}ay private
network is located in the EMBRACE container.  It is also on another
dedicated network and controls a number of subsystems including power
supplies, local oscillator (LO), as well as the thermometer readout.
The same computer handles the control of the ADC boards through USB
communication with the pointing controllers of the 4 antennas.

\begin{figure}
\centering
\includegraphics[width=0.8\columnwidth]{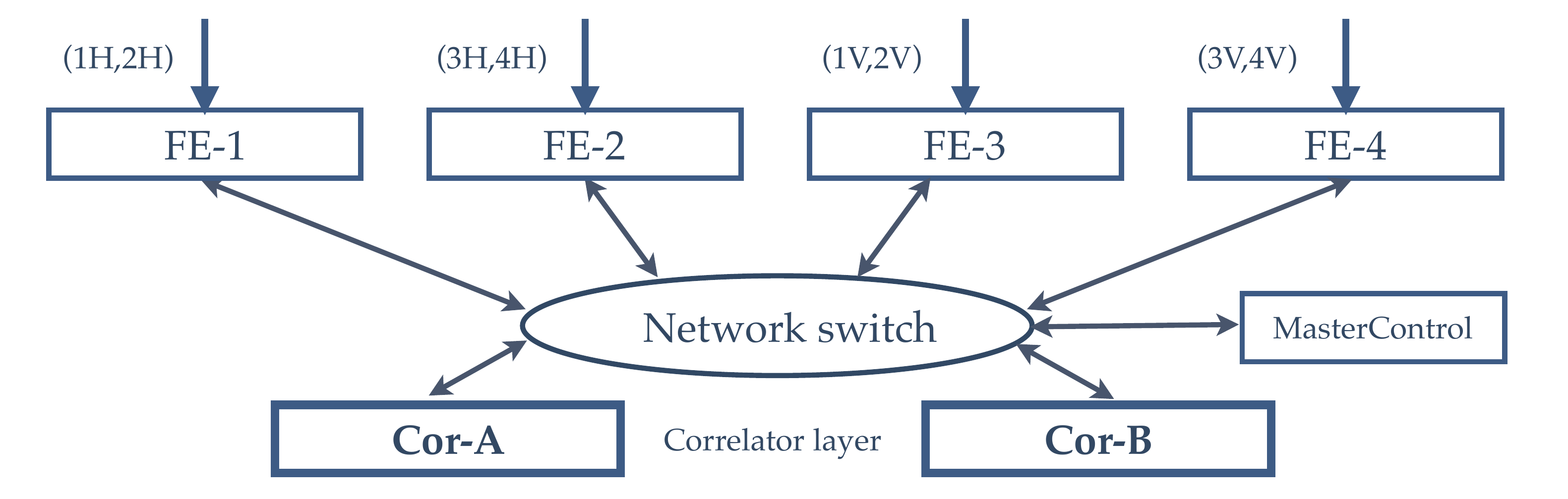} \\[1mm]
\caption{Schematic view of the PAON4 acquisition and correlator system } 
\label{Fig:acqsystem}
\end{figure}

The acquisition/software correlator system itself is composed of
6~computing nodes and an additional machine acting as the acquisition
controller and supervisor.  The computing nodes are arranged in two
layers: 4 of them form the first layer (front-end machines) which
receive data from the two ADC boards, through four optical links. The
two nodes forming the second layer perform the correlation
computations. Figure~\ref{Fig:acqsystem} shows a schematic view of the
system configuration.
The acquisition and processing of the two ancillary RFIMON and THERMON
signals is done by an independent computer running the same software.

The front-end machines perform FFT on the received data streams,
except when the FFT firmware is used, and then divide the frequency bands
into a number of sub-bands corresponding to the number of computing
nodes in the correlator layer.  Each front-end machine receives the
data corresponding to two RF signals, currently arranged as a pair of
polarization signals, H-H or V-V.  Fourier components for each
sub-band is sent to the corresponding compute node in the correlator
layer which has two nodes, each computing 36 correlations
for 2048 frequency channels over 125~MHz band which is half
of the full ADC board Nyquist sampling limit of 250~MHz.  The 36
computed correlations correspond to the following pairs:
\begin{itemize}
\item 8 auto-correlations, 4 corresponding to the H signals, and 4 to the V signals.
\item 12 correlations corresponding to the cross correlations between the same polarization signals. There are 6 H-H and 6 V-V  cross-correlations between H-polar and  V-polar signals from the 4 feeds. 
\item 16 correlations corresponding to H-V cross-correlations between the H-polarization signal from one feed and 
the V-polarization signal from the second feed. 
\end{itemize}

The total data rate ingested by the front-end machines at $\sim 10 \%$
on-sky time corresponds to $\sim 400 \mathrm{MB/s}$, with a factor 4
higher rate exchanged between the two layers, as the FFT computation
is carried with 4-byte floating point numbers.  Correlations are taken
typically at a sampling rate of one or a few seconds. Most data
discussed here was taken with a correlation sampling rate of 3 or 6
seconds. Computed correlations (also called ``visibilities'') are
saved into permanent storage (disk) every few seconds, and a number of
visibility matrices, defined again by input parameters are grouped
into a single file.
A typical 24~hours observation run with 3~second visibility sampling rate produces around 70~GB of data, 
split into $\sim 100 \mathrm{MB}$ size files. 
 
The 
data acquisition and correlator software, \textbf{TAcq}, was developed
for the BAORadio project starting from 2007 and has been enhanced
over the years. It is a flexible, high performance C++ software
package, which uses the SOPHYA\footnote{ SOPHYA C++ class library :
  \tt http://www.sophya.org/ } class library for standard numerical
algorithms and I/O.  The flexibility and various operation
modes allows \textbf{TAcq} to be used at different stages of the project, from final
tests of the electronic system boards, to the routine operation of the
interferometer, as well as during commissioning.

The \textbf{TAcq} package defines a simple and very efficient
multi-threaded architecture, where different threads communicate through
a central memory manager hub.  The number of running threads is also
dynamically determined through the parameters which are tuned
to a given hardware configuration.  A description of the software
architecture, the different classes and system operation modes are
beyond the scope of this paper and will be discussed in a separate,
forthcoming publication.

In practice, a main program called \textbf{mfacq} with different
operation modes, controlled by an extended set of parameters handles
the acquisition and correlation computation tasks.  The control node
spawns one \textbf{mfacq} process in each of the compute nodes, the
four front-end nodes and the two nodes in the correlator layer for
PAON4, and provides each process their specific set of parameters. The
complete system operates in streaming mode, with data packets
exchanged between \textbf{mfacq} processes through the network
layer. The time synchronization for visibility computation is ensured
by the hardware time tags which are propagated through the data
packet streaming protocol.

\begin{figure*}
\centering
\includegraphics[width=0.7\textwidth]{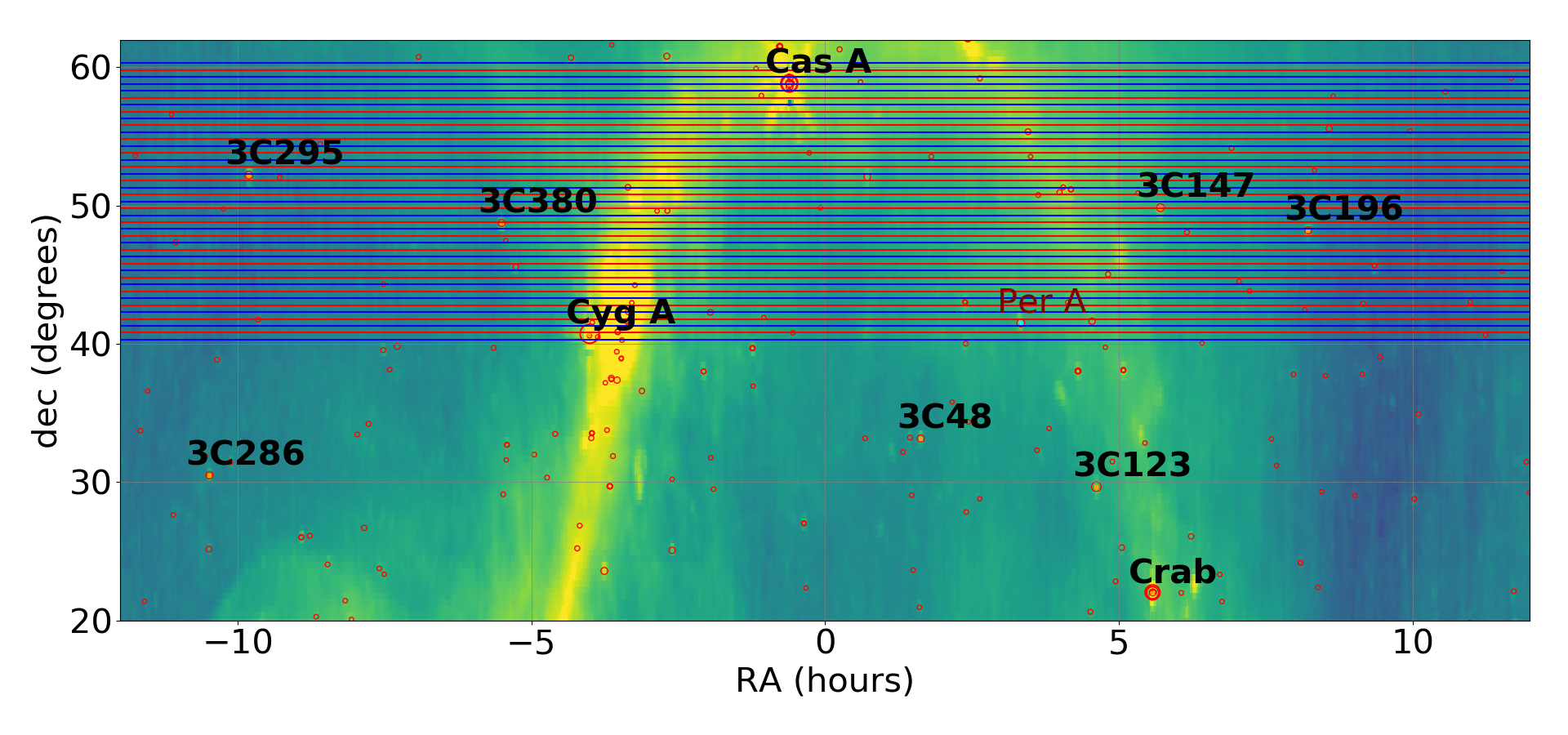}
\caption{PAON4 acquisitions in summer and fall of 2018, in the
  (RA,dec) sphere with equatorial projection.  Each acquisition is
  represented by an horizontal line, blue or red for the two halves
  (see text). The position of bright sources from NVSS (with the
  addition of Perseus~A) is shown by open circles.  The diameter of
  the circle represents the source brightness.  The underlying image
  is a projection of the Haslam 408~MHz synchrotron map, prior to
  source subtraction, as retrieved from the LAMBDA archive }
\label{Fig:data_scan_2018}
\end{figure*}
 
\section{Operations and observations}
\label{Sec:Observations}

After the completion of its construction (mid 2015), the PAON4
observations served several aims in successive periods, with few minor
hardware or software changes. Observations carried out in 2015 were
mostly dedicated to hardware debugging, from finding faulty cables and
electronic modules to identifying some RFI sources. For example, an
Ethernet switch was initially installed in the central antenna
electronic cabinet, to ensure the readout of a temperature sensor, and
was generating RFI. We also spent some time measuring correlated
noise, both on the instrument and in the lab, in 2015 and 2016.
Correlated noise measurements in the lab showed that the main source
was by coupling through the power supply. The overall system was
subsequently improved with better shielding, more careful ground
wiring, and an improved power supply system. The correlated noise
level was reduced by a factor~10 in this process. Gain variations and
some randomly occurring perturbations were observed. Clarifying their
sources and curing or correcting these effects consumed significant
time and effort in 2017 and early 2018, as discussed in
Section~\ref{Sec:Data-processing}.

In May 2016, the 4V channel was terminated by connecting its front-end
LNA to a 50$\Omega$ resistor.  This was done as a diagnostic for
understanding the source of perturbations (see
Section~\ref{Sec:external-perturbations}) and gain variations (see
Section~\ref{Sec:spectral-response-time-gain}).  The 4V channel was
reconnected antenna and the additional THERMON signal was added to the system in
July 2018, after the identification of the perturbation source. In
2015 and 2016, we were operating PAON4 with the FFT firmware loaded on
the ADC boards.  The ADC boards were then sending FFT coefficients to
the software correlator. We switched to the RAW firmware in February
2017, in order to investigate some problems with the FFT firmware,
including possible saturation by satellite signals because of the
limited dynamic range of the FFT firmware.  We also added two front-end
nodes to handle FFT computation by the acquisition/correlator
software.
\begin{figure*}
\centering
\hspace*{-7mm}
\begin{tabular}{cc}
\setlength{\tabcolsep}{0pt}
\hspace*{-5mm}
\includegraphics[width=0.53\textwidth]{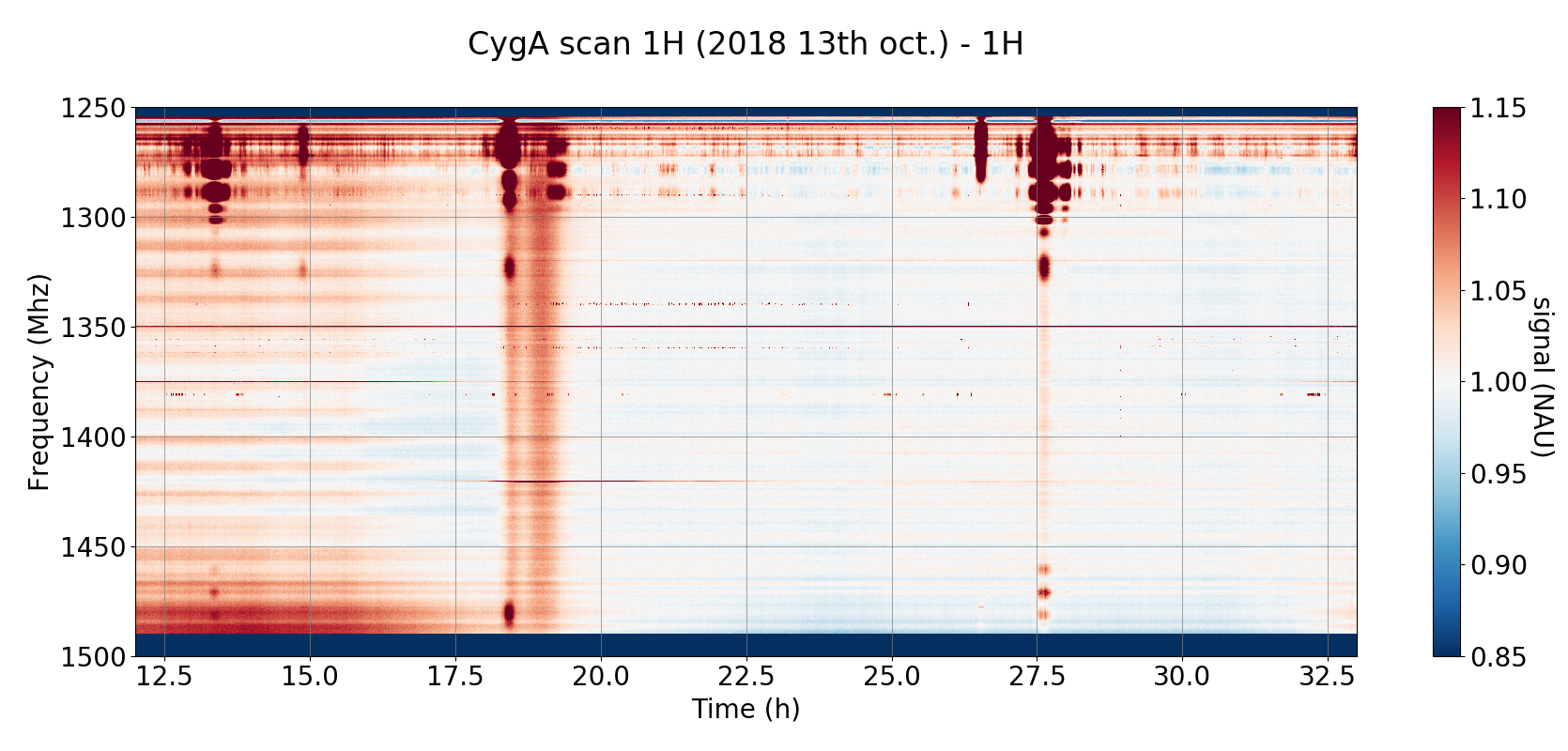} &
\hspace*{-5mm} \includegraphics[width=0.53\textwidth]{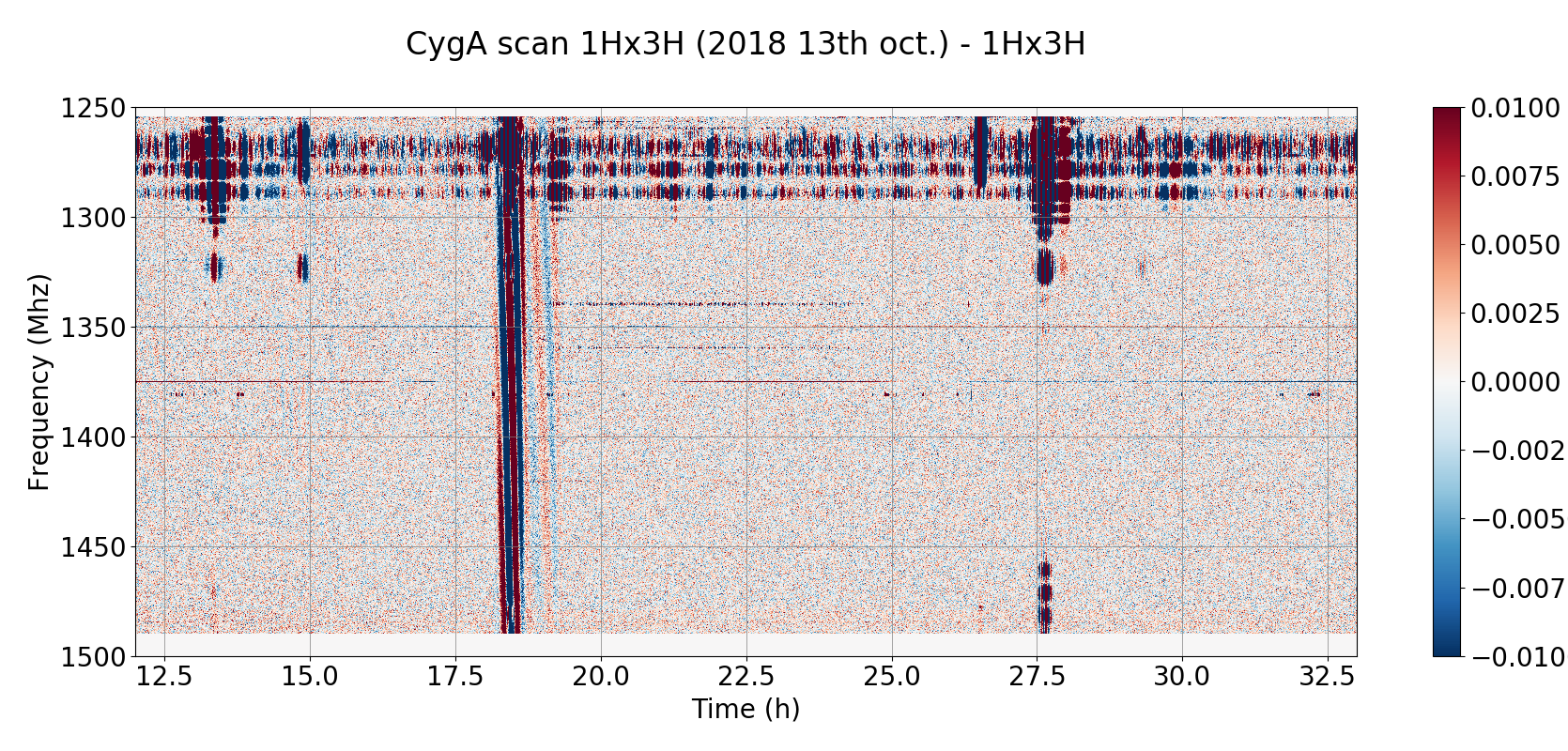} 
\end{tabular}
\centering\includegraphics[width=0.5\textwidth]{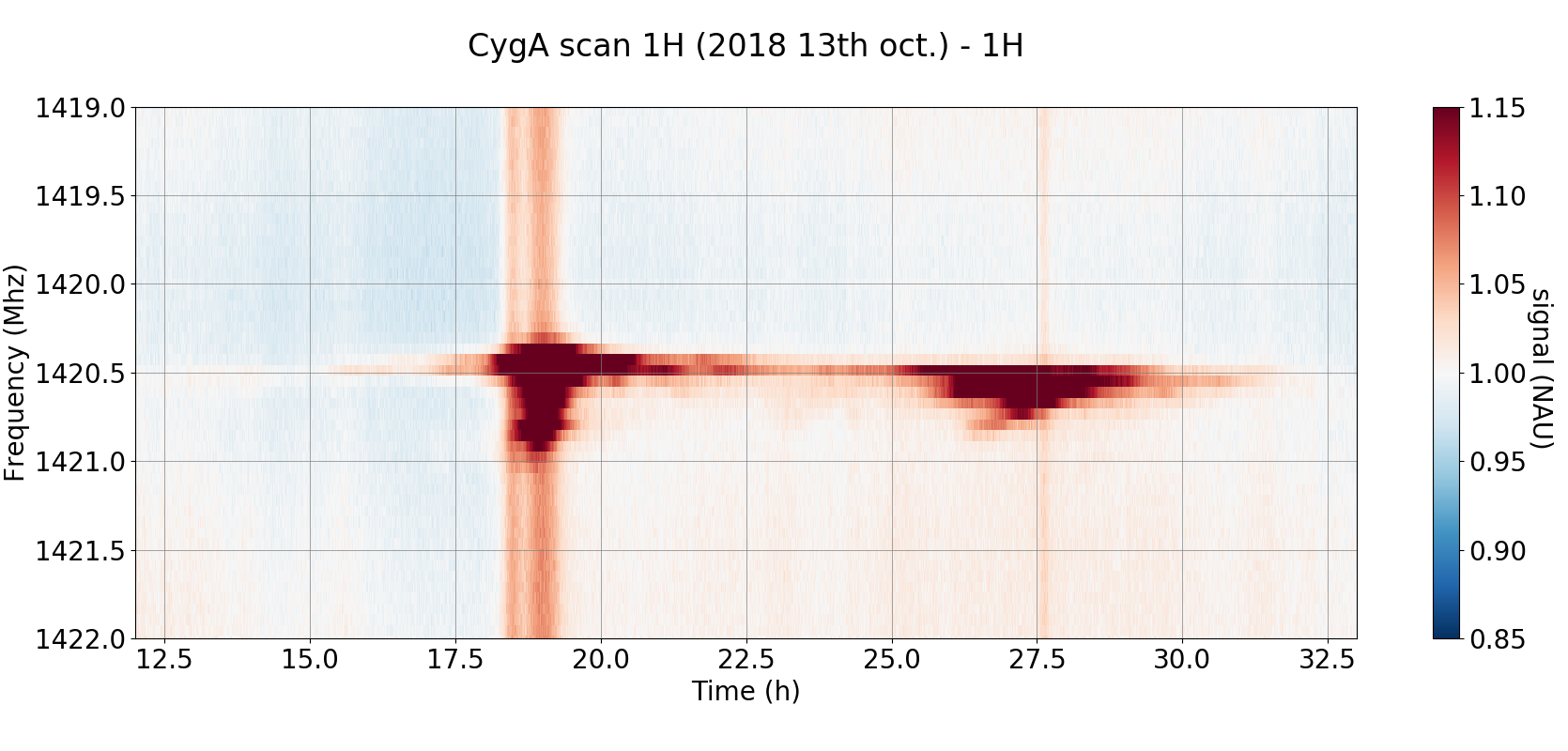}
\caption{\textbf{Top row:}~~Examples of time frequency maps, in
  normalized units (NAU) obtained from processing PAON4 visibility
  data (after gain correction).  Time (UT) is along the horizontal
  axis, spanning about 24 hours, starting from 0h UT on the starting
  day (Oct. 13th in this example). The vertical axis corresponds to
  frequency, varying from 1250~MHz to 1500~MHz, from top to bottom.
  The autocorrelation signal 1H is shown on the left, while the cross
  correlation signal $1\mathrm{H}\times3\mathrm{H}$ is shown on the
  right. The transit of Cyg~A can be seen at about 18h30~UT followed
  by bright galactic emission.  Several satellite transits are also
  clearly visible, including one just before the Cyg~A transit.
  Residual RFI and time variation of the stationary wave pattern in
  $g(\nu)$ before Cyg~A transit in the autocorrelation map are also
  visible.  \textbf{Bottom row:}~~zoom of the autocorrelation of the
  1H channel arround the 21~cm frequency. }
\label{Fig:tfm_auto_cross}
\end{figure*}

Data taking is automatic once all acquisition parameters are set, and
can be remotely initiated and monitored.  After each acquisition run,
data are transferred to the IN2P3 computing center for analysis and
archival.  An SSH connection to the acquisition cluster is used for
launching the data taking.  It sets up the antenna and configures the
digitization boards, and checks or modifies the acquisition parameters
such as the visibility sampling rate, the observation ID, and the data
set path on the master control node. The acquisition is started or
scheduled to begin at a later time.

PAON4 instrument has been designed to survey a significant fraction of the northern sky, 3000-5000 $\mathrm{deg}^2$,
with full right ascension coverage, through a set of 24~hour constant declination scans. 
In fall 2016, we carried three mini surveys, observing declination
ranges around the Cygnus~A and Cassiopeia~A bright radio sources with
the FFT firmware.  Details can be found in Table~\ref{Tab:listeobs}
and a more in-depth analysis can be found in \citep{2019PhDQHuang}.
As mentioned above, observations from beginning 2017 until mid 2018
were motivated by the necessity to identify the source of
perturbations.  We started the first extended PAON4 survey in July
2018, covering about 20~degrees in declination, between the
declinations of Cas~A and Cyg~A.  More than forty 24-hour constant
declination scans with 0.5~degree step were done from July to December
2018.  The observations were split into two interleaved independent
sets, labelled Scan\_2018\_A and Scan\_2018\_B. In each set, scans are
separated by 1~degree step and the scans in B are shifted by
0.5~degrees with respect to the corresponding A~scan.  The scan
pattern for Scan\_2018 (A \& B) data sets is shown in
Figure~\ref{Fig:data_scan_2018}.  A number of additional problems have
been identified by analysis of the Scan\_2018 data, in particular
imperfect packet synchronization by the ADC boards in some cases. The
2019 observation strategy was then targeted to further investigate
these problems and to precisely characterise the instrument (pointing,
lobe \ldots) through long duration (LD), uninterrupted observations
over several days. Table~\ref{Tab:listeobs} summarises the
characteristics of the different PAON4 data sets.
\begin{table*}
  \begin{tabular}{c c c c c c }
   \tableline
            DataSet & Period      &  declination (deg) &  Nb.Scans  &  Duration  &  comment \\
    \tableline
           CygA-Sep16 & Fall 2016 & 37...44  & 10 & 2 weeks & FFT firmware \\
           CasA-Sep16 & Fall 2016 & 56...60  & 5 & 2 weeks & FFT firmware \\
           CygA-Nov16 & Fall 2016 & 35...46  & 11 & 2 weeks  & FFT firmware \\
           Scan\_2018 A & Summer 2018  & 40...60 & 20  & $\sim 2$ months & $\delta_{step} = 1 \mathrm{deg}$ \\
           Scan\_2018 B & Fall 2018  & 40...60 & 20 & $\sim 2$ months &  $\delta_{step} = 1 \mathrm{deg}$  \\
           LD\_1 & January 2019 & 25, 30, 42, 58 & 4 & $\sim 10$ days  & CasA, CygA, M1, VirA  \\
           LD\_2 & April/May 2019 & 12...24 & 7 & 2 weeks  &  M1, Sun \ldots \\ 
           LD\_3 & June,July 2019 & 35  & 1 & 10 days & CygA, stability checks \\ 
           LD\_4 & July 2019 & 12 , 22, 58 & 3 & $\gtrsim 2$ weeks  & CasA, M1, VirA \\ 
    \tableline
   \end{tabular}
\caption{PAON4 data sets  }
\label{Tab:listeobs}
\end{table*}



\section{Data processing}
\label{Sec:Data-processing}

The data processing pipeline is composed of a set of C++ programs and
python scripts which perform the tasks listed below.  The visibility files are the input of the
processing pipeline.  They contain
$N_\mathrm{vis} \times N_\mathrm{freq}$ matrices, one matrix per
visibility sampling time, where $N_\mathrm{vis}$ is the number of
visibilities, and $N_\mathrm{freq}$, the number of frequency channels,
which are respectively $N_\mathrm{vis}=36$ and
$N_\mathrm{freq}=4096$ for PAON4.  Most of the data analyzed here has
a visibility sampling rate of 3~s or 6~s.  The pipeline includes
the following components:
\begin{enumerate}
\item data quality monitoring and noise characterization.
\item Determination and correction for time dependent electronic gain
  variations $G(t)$ as well as frequency dependent gain $g(\nu)$.
\item Production of time-frequency binned (TFM) or right ascension-frequency
  binned (RAFM) maps for the different visibilities.
\item Phase and gain calibration using known sky sources, and
  determination of instrument parameters such as system temperature
  $\Tsys$ and beam response.
\item RFI flagging and cleaning, mostly performed on TFM maps.
\item Production of fully cleaned and calibrated time-frequency binned.
  (TFM) or ascension-frequency binned (RAFM) maps for the different
  visibilities which are then used by the subsequent analysis stages,
  in particular by the map making.
\item Map making through different methods, including m-mode
  decomposition \citep[see e.g.][]{2016MNRAS.461.1950Z}
\end{enumerate}

The mathematical formalism for radio interferometer measurement
equations and the calibration methods have been studied by many
authors, including polarimetry issues
\citep[e.g.][]{1996A&AS..117..137H,1996A&AS..117..149S,2011A&A...527A.106S}.
The numerical optimization problem has also been studied
\citep[e.g.][]{2014A&A...571A..97S, 10.1093/mnras/stv418} as well as
their application to the calibration of large arrays such as LOFAR,
MWA or HERA
\citep{2011MNRAS.414.1656K,2016ApJS..223....2V,2019ApJ...875...70B}.
However, specificities of mid frequency arrays operating in transit
mode have not yet been fully investigated.

In this paper, we focus on the first four stages of the pipeline,
listed above. The discussion of subsequent stages and the maps
produced from PAON4 observations is beyond the scope of this paper,
and will be the subject of forthcoming publications.

Typical time frequency maps (TFM) spanning 24~hours, obtained after a
basic processing of PAON4 visibility data are shown in
Figure~\ref{Fig:tfm_auto_cross}.  These maps were obtained from data
taken on 2018~October~13, with antennas pointed toward the Cyg~A
declination at $\delta=42^\circ$.  The left panel shows the
auto-correlation signal from the 1H channel.  The strong pollution of
the lower part of the frequency band, below $\sim1300\,\mathrm{MHz}$
by RFI, mostly from satellites, is clearly visible.  The transit of
bright sources, here Cyg~A and the galactic plane, can be seen on
the autocorrelation map.  The galactic 21~cm emission around 1420~MHz
is also visible, with the zoom around this frequency in the lower
panel showing the galactic frequency dependent emission pattern.
The right panel shows the cross-correlation signal from the (1Hx3H)
feed pair.  In addition to gain correction, an average frequency
template corresponding to the correlated noise has been subtracted
from the cross-correlation time-frequency map. The transit of the
bright source is clearly visible with high signal to noise.  Satellites are also visible in the lower frequency range.
\begin{figure}
\centering 
\setlength{\tabcolsep}{0pt}
\begin{tabular}{c}
\includegraphics[width=0.99\columnwidth]{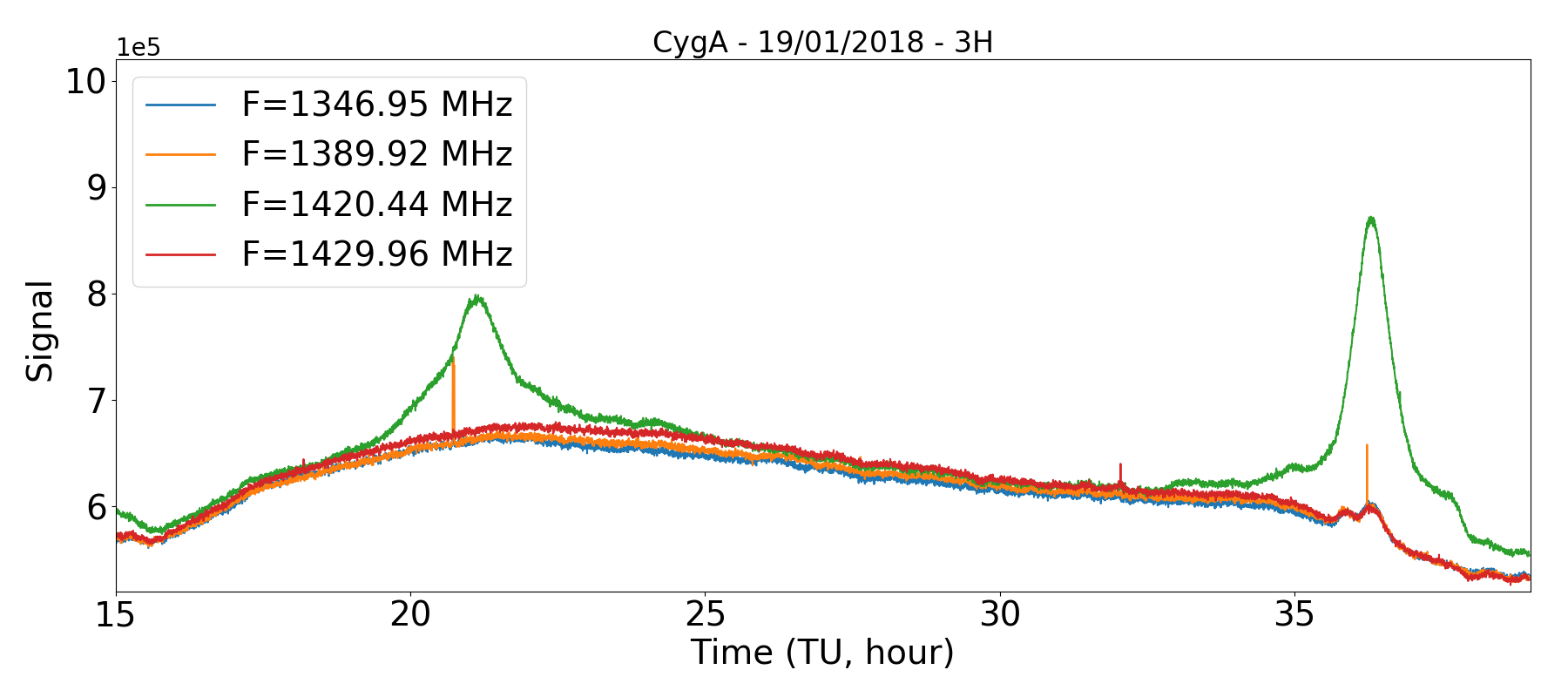}\\
\includegraphics[width=0.99\columnwidth]{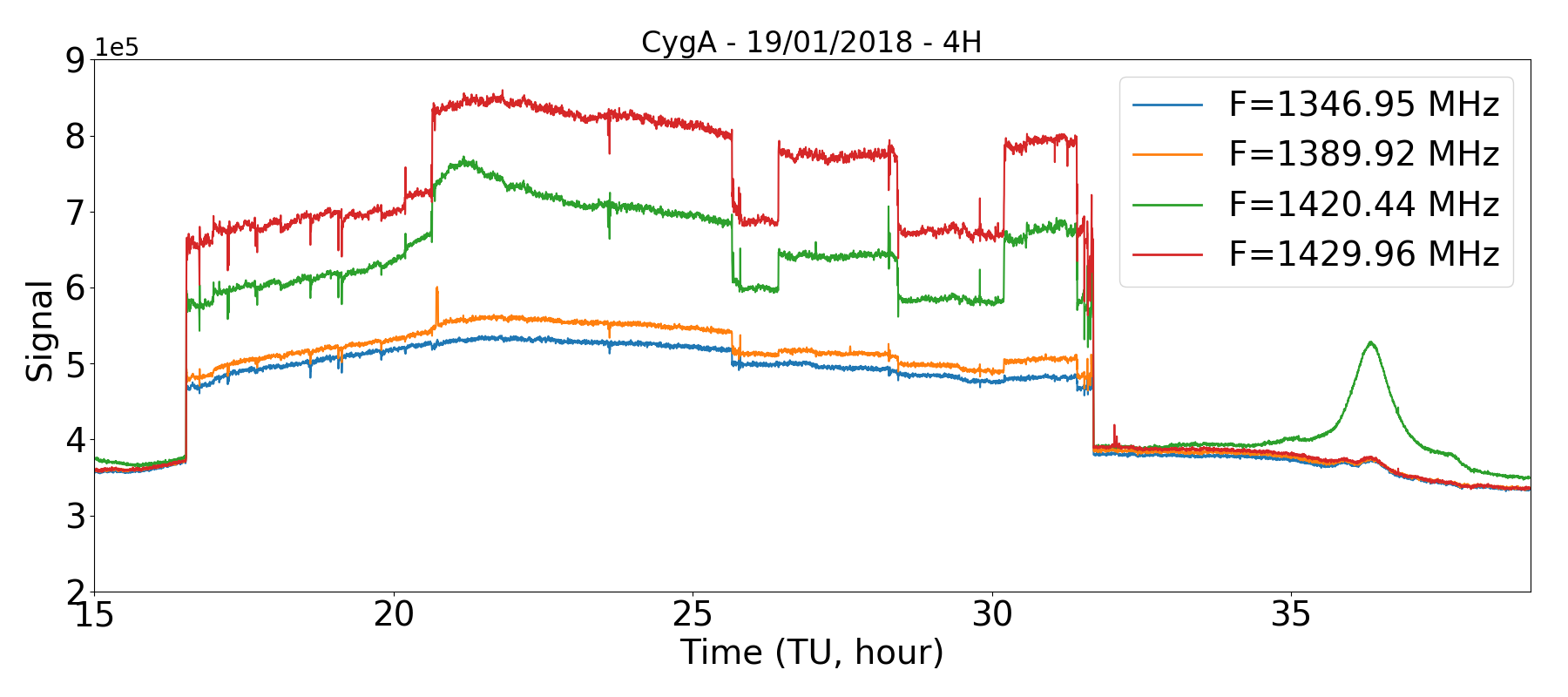}
\end{tabular}
\caption{Example of variations of autocorrelation levels as a function
  of time measured on PAON4 for two channels (\textbf{top:}~~3H, \textbf{bottom:}~~4H)
  on January 19th, 2018, at 4 frequencies (\textbf{blue:}~~1347~MHz, \textbf{orange:}~~1390~MHz, \textbf{green:}~~1420.5~MHz,
  \textbf{red:}~~1430~MHz) over a duration of
  about 24h. Antennas were set at the elevation of the transit of Cygnus~A.
  During this observation, channel 4H was perturbed.}
\label{Fig:perturbation}
\end{figure}

\subsection{External perturbations}  
\label{Sec:external-perturbations}

Before July 2018, PAON4 data were perturbed by a persistent artefact.
Finding its origin occupied a major fraction of PAON4 observing time
in 2017 as well as during the first semester of 2018.  Sudden changes
in the signal level were observed on a few of PAON4 signals.
Figure~\ref{Fig:perturbation} shows the variations of signal levels at
four frequencies, for the 3H signal on the left, with normal
behaviour, and the 4H signal on the right, exhibiting sudden level
changes or jumps, at all four frequencies. On the normal 3H
autocorrelation signal, one may observe a slowly varying trend, which
is due to temperature induced gain variations, following day-night
temperature changes.  The transit of Cygnus~A, at about 36~h
(i.e. 12:00~UT), is visible on all frequencies except at 1420.44~MHz
where 21~cm galactic emission dominates.  In contrast with this normal
behaviour, the perturbed 4H signal exhibits erratic variations,
starting from a sharp rise at about 17~h and ending after a sharp
fall at about 32~h (8:00~UT next day), with several intermediate positive
or negative steps, at all frequencies.  These perturbations were already
seen on early PAON4 data, in 2015, but were more frequent during the
2017-2018 year. Investigating this effect during that year, we noted
that:
\begin{itemize}
\item[-] The perturbation occurrence rate was $\sim80$\% (i.e. 8 days out of 10)
\item[-] Only one antenna was affected at a time
\item[-] The most affected antenna was antenna 4, with a rate of about
  80\%.  However, in 2015, whenever the effect was observed it
  affected only antenna 3
\item[-] In the rare occurrences where one of the another antennas was
  affected, both polarizations were affected by similar (but not
  identical) perturbations, in exactly the same time interval.  In the
  case of antenna~4, one of the two polarization probes on the feed,
  the 4V channel, was terminated from mid 2016 to mid 2018 by a
  $50\,\Omega$ resistor connected to the LNA input. No significant
  perturbation was ever observed on this terminated channel.
\end{itemize}

\begin{figure}
\centering
\setlength{\tabcolsep}{0pt}
\begin{tabular}{ll}
\includegraphics[width=0.8\columnwidth]{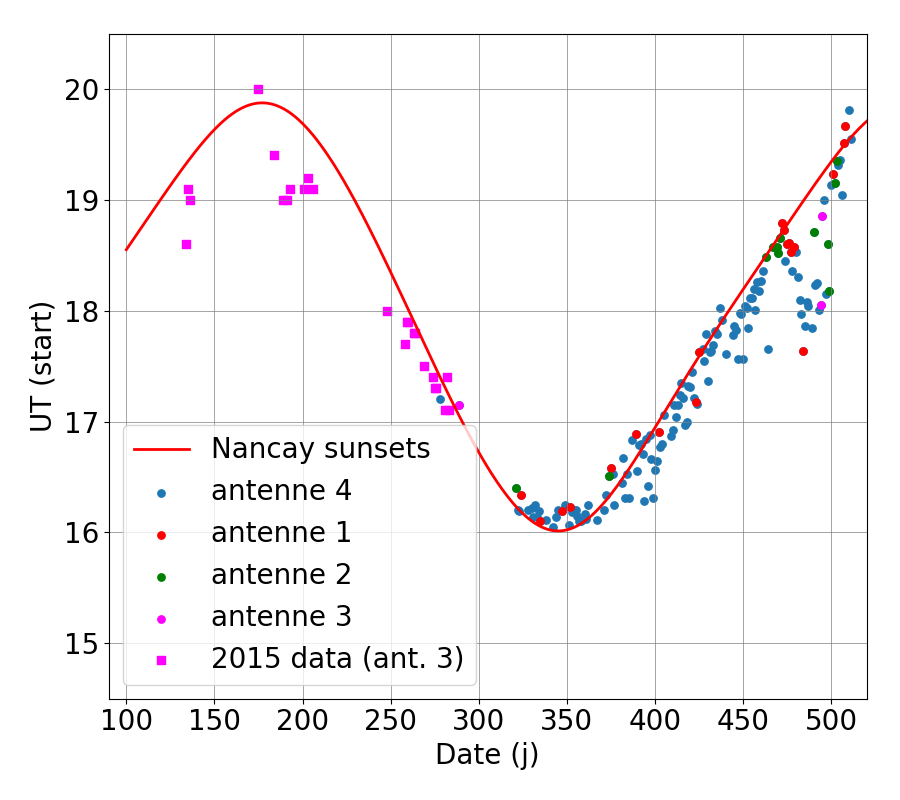}\\
\includegraphics[width=0.8\columnwidth]{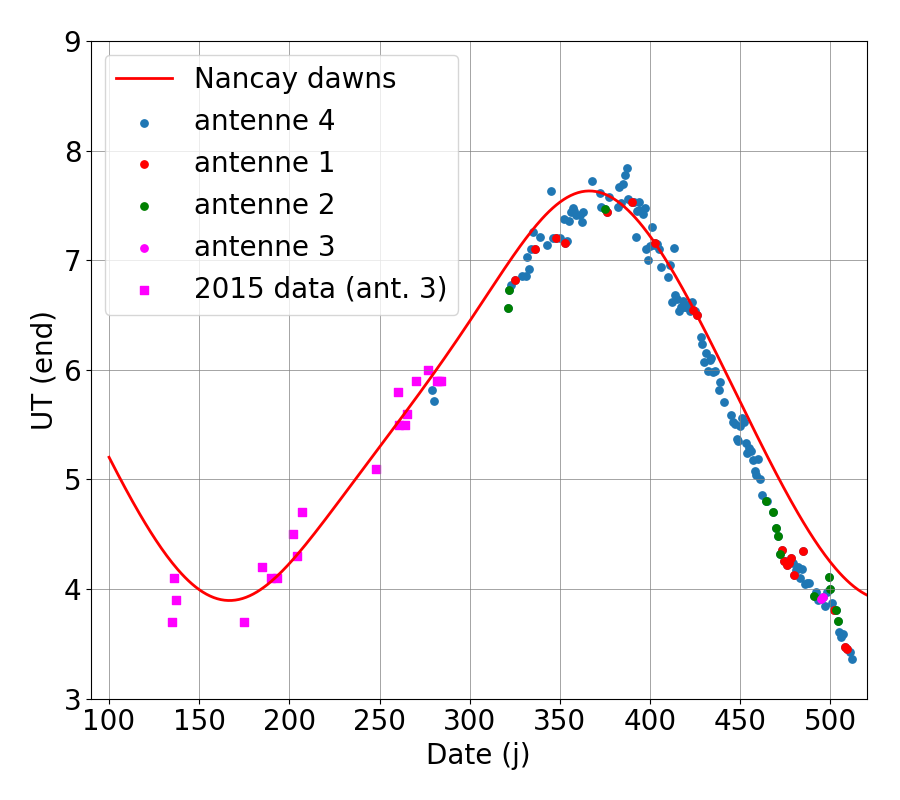}
\end{tabular}
\caption{Variation of the start time (top) and end time (bottom) of
  the perturbation with date. The abscissa shows the date,
  reported as a number of days since January 1st, and the ordinate
  corresponds to the time in hours.  The colour of the symbols
  corresponds to the antenna which was perturbed at each date. Early
  PAON4 data, in April-May 2015, when only antenna~3 was perturbed, is
  shown as squares.}
\label{Fig:perturbation_timing}
\end{figure}

 This last fact, combined with other checks through electronic modules
 and cable permutations showed that the perturbation had an external
 origin.  Accumulating more data during the year brought another
 intriguing piece of evidence, displayed in
 Figure~\ref{Fig:perturbation_timing}. The start and end times of the
 perturbation seemed to correlate quite well with the hour of the dawn
 and sunset in Nan\c cay, although with some dispersion.  After a long
 investigation campaign, we finally spotted in May 2018 a small bird
 spending its nights in the shielded environment of the chokes
 surrounding our feed.  We concluded that the thermal radiation
 produced by the bird was the source of an additional signal during
 the night.  We placed anti-bird nets around the chokes end of May
 2018 which cured the problem.

\subsection{Spectral response and time dependent gain}
\label{Sec:spectral-response-time-gain}

The PAON4 channel response varies both with time and frequency. We use
in our analysis a simplified gain correction model by considering
these two effects independently of each other, giving
$G(t,\nu)=G(t)\times g(\nu)$.

\begin{figure}
\centering
 \includegraphics[width=0.8\columnwidth]{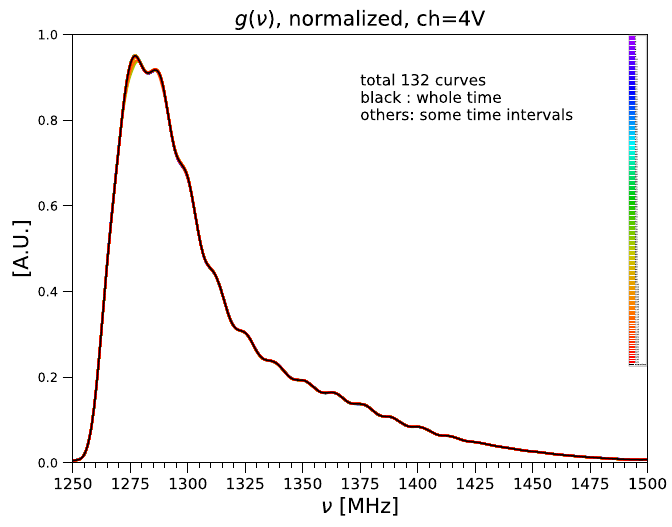} \\
\includegraphics[width=0.8\columnwidth]{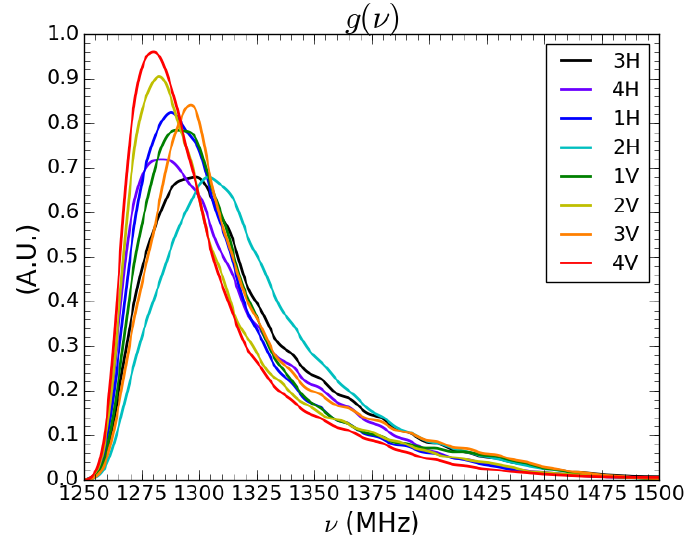} 
\caption{Frequency dependent gain determination from November 2016
  data. \textbf{Top:}~~Stability of the frequency dependent gain
  $g(\nu)$ over time. The 4V channel gain is shown, with the total
  time interval split into 131~intervals. The black curve shows the
  $g(\nu)$ for the whole period, while gains for the 131~intervals are
  shown in different colors.  \textbf{Bottom:}~~The normalized
  frequency dependent gains $g(\nu)$ for the eight PAON4 RF signals.}
\label{Fig:gain-g-nu}
\end{figure}

As a result of the combination of the various stages of frequency
filtering, and signal transmission cables, the frequency response is
complex (Figure~\ref{Fig:gain-g-nu}).  The gain $g(\nu)$ shape is
primarily determined by the LNA and pre-amplifier spectral response,
but in the current design of the DAQ, the 50~m coax cables between the
central dish and the digitization boards located in the EMBRACE
container are responsible for the dramatic gain decrease above
1300~MHz.  For each channel, the gain also shows wiggles that are
attributed to standing waves in the 9~m cable between the LNA and the
pre-amplifier and the 10~m cables between the pre-amplifier and the
local oscillator electronics.  The characteristics of these standing
waves are time variable.  The peak-to-peak wave separation, of the
order of 12.5~MHz, is inversely proportional to the cable length which
motivated the shortening of cables as much as possible.  The R\&D
program began at this time with the aim of performing the analogue to
digital conversion as close as possible to the LNA on the feed.  This
R\&D program will result in the installation of new electronic boards
called IDROGEN (see Section~\ref{Sec:Futur}).

As mentioned above in
Section~\ref{Sec:external-perturbations}, and shown in
Figure~\ref{Fig:perturbation} (right hand side), the raw PAON4
auto-correlation signals in each channel show a systematic
drift in time with a day-night pattern.  This is caused by
the gain variations due to ambient temperature variations experienced
by the analogue electronics chain.  The analogue electronics are
located outside without any environmental control system.  The
temperature stability of the PAON4 electronics chain was measured
in the lab.  There is 4~dB ($\sim 60\%$) gain variation for
temperature varying between $0^{\circ} C$ and $55^{\circ} C$.  In the
PAON4 setup, a $15^{\circ} C$ range is expected, and with the sun
shine effect, the daily amplitude of the temperature variation could
easily reach and exceed $30^{\circ} C$ leading to $\sim 30\%$ gain
variations. {\color{black} The analogue electronics were not designed to
  be deployed without a temperature-controlled environment and this is
  an issue which will be improved in the next generation electronics
  design (see Section~\ref{Sec:Futur}).}

Using data taken in 2017 and 2018 with the resistor terminated 4V
channel, we demonstrated that time variation of the signal of each
channel, outside bright sky source transits times, was tightly
correlated with the terminated channel.  A gain variation model based
on this correlation was then created and the terminated channel signal
used to compute and correct for the time varying gain term $G_i(t)$ of
all other channels.  The effect of gain variations decreased
from $\sim25-30\%$ to $\pm3-5\%$ using this simple scheme.  The~4V
channel was reconnected in July 2018, and an additional resistor
terminated channel, THERMON, was added to PAON4.  This is now used to
correct for gain variations.  The gain corrected visibilities are
expressed in NAU {\it (Normalized Arbitrary Unit)}, with
auto-correlation levels close to unity outside bright sources or sky
transits.

\begin{figure}
\centering
\setlength{\tabcolsep}{0pt}
\begin{tabular}{cc}
\includegraphics[width=.8\columnwidth]{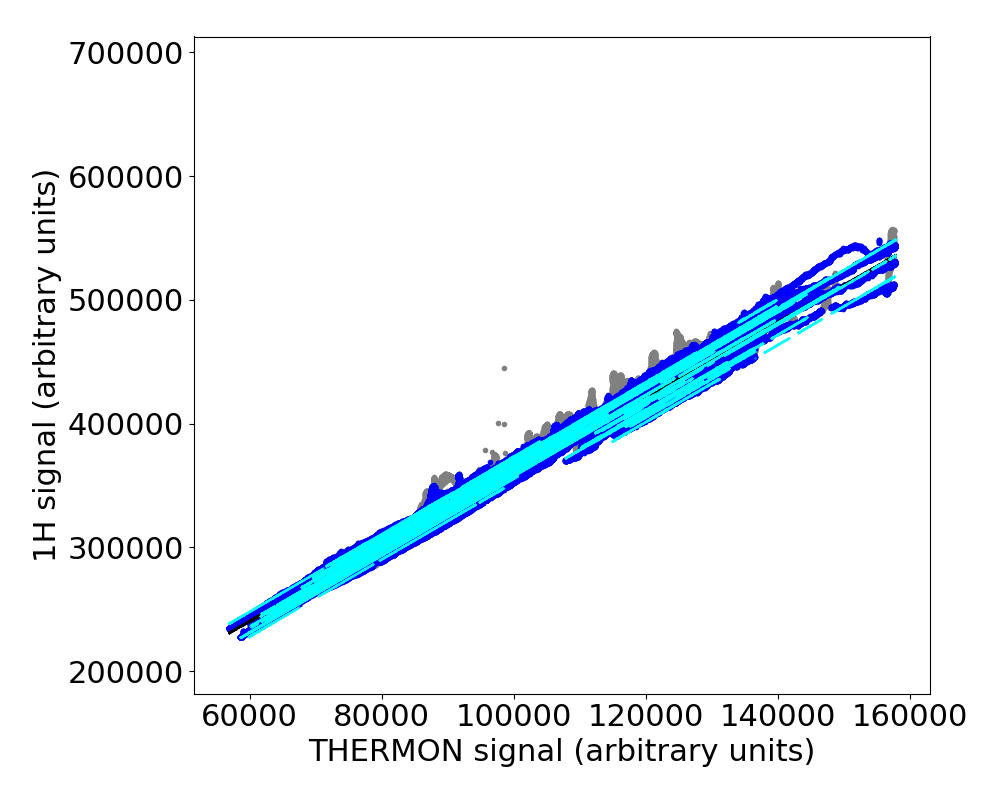}\\
\includegraphics[width=.8\columnwidth]{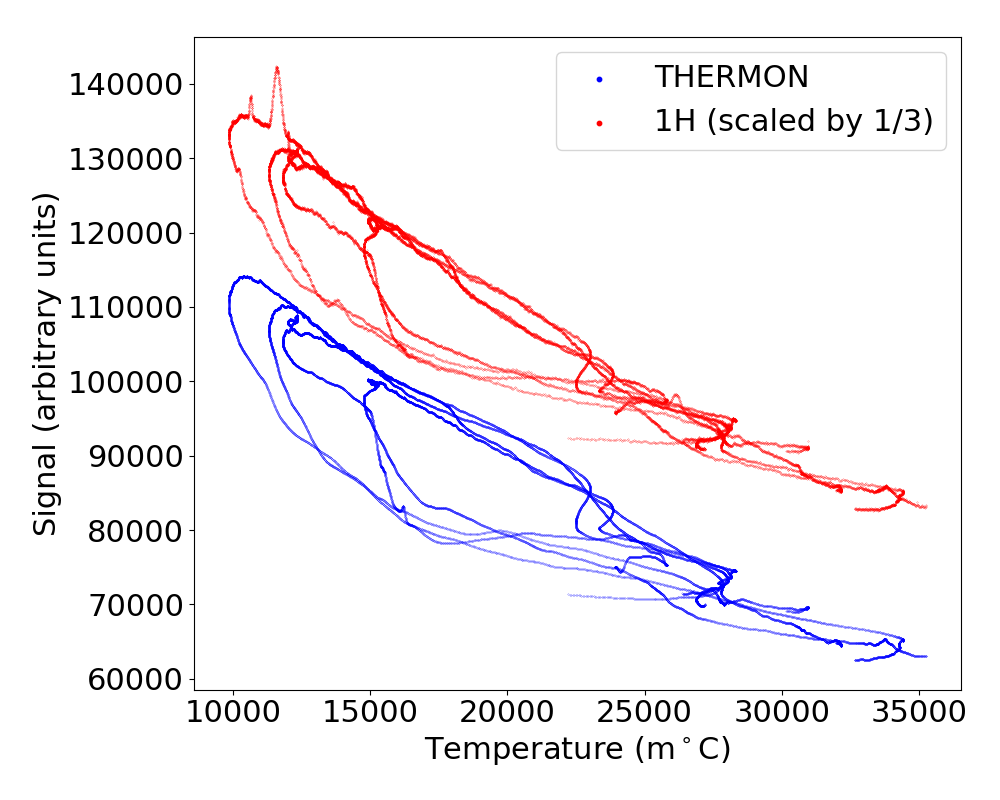}
\end{tabular}
\caption{\textbf{Top:}~~Time dependent gain $G(t)$ correction, for
  channels 1H, for good data from Scan\_2018 A and B. Gray dots are
  excluded from the fits (due to source transit, satellites, or RFI).
  Only the blue ones are used.  The cyan thin lines represent the
  results of the global fit, with one slope for all datasets and one
  intercept for each of the observation runs.
  \textbf{Bottom:}~~Variations of the THERMON and 1H channels as a
  function of the measured ambient temperature over a few days in July
  2018.}
\label{Fig:gain-G-t}
\end{figure}

We determined for each channel the parameter of a linear correlation
with the median of the signal (after satellite and source removal) and
the THERMON signal (in the same time frame).  We show in
Figure~\ref{Fig:gain-G-t} examples of this reconstruction for the
1H~channel when analyzing data from the Scan\_2018 A and B part
simultaneously.  A unique slope was fitted with different intercepts
for each observation run.  Once correcting for the linear variations,
outlined in cyan, the signal variation amplitudes are reduced from
$\sim25-30\%$ to $\pm3-5\%$.  The THERMON and the antenna channels are
both correlated with a direct temperature measurement.  This
correlation has a large dispersion and shows hysteresis as a result of
the different mechanical structures and the positioning of the
temperature sensor (on top of the EMBRACE hut).
Figure~\ref{Fig:gain-G-t} shows that the THERMON signal provides a
better template for correcting the temperature-induced signal
variations than a direct temperature measurement.  Evidence of a
residual hysteresis may be seen on the left panel, which could be
explained by the thermo-mechanical differences between the LNA's on
the feeds and the lighter THERMON set-up, leading to faster
temperature variations for the latter.
 
\subsection{On sky calibration}
\label{Sec:on-sky-calibration}
Point sources on sky are used to determine the instrumental phases, as
well as the antenna beam and overall radiometric calibration.  The
PAON4 survey strategy includes regular observation of bright sky
sources, such as Cassiopeia~A (Cas~A) and Cygnus~A (Cyg~A), every few
days to obtain gain and phase calibration parameters which are then
used for the subsequent scans and to update the instrument model.

The primary calibration source is Cas~A which can be considered a
point source for PAON4.  A simple model is used to represent expected
auto-correlation and cross correlation signals from Cas~A, and we fit
instrumental phases, amplitudes, as well as beam shape parameters.
Also fitted is the effect on the auto- and cross-polarizations for
H~and~V due to changing the pointing direction along the East-West
direction.  For each polarization, H or V, independently from each
other, the parameters are adjusted on the set of 4~auto-correlations
and 6~cross-correlations signals, for each binned frequency channel.
The cross-correlations between two orthogonal probes H$\times$V are
not included, ignoring source polarization.  After fitting the
auto-correlations which notably fix the beam shape parameters, the
6~cross-correlations signals are used to determine relative phases
between the feed on antenna~1 taken as reference and the 3 other feeds
(i.e $\Delta\Phi_i=\Phi_i-\Phi_1$ with $i=2,3,4$).  There are
degeneracies between the instrumental phases and the relative phases
induced by the array geometry, which is fixed at this stage.  We have
refined PAON4 antenna positions, determined from geometrical
measurements during PAON4 deployment, using satellite transits as
described below.

The lower part of PAON4 band, below $1300\,\mathrm{MHz}$ is heavily
polluted by satellite emissions, dominated by positioning satellites
from the GPS (US), Galileo (EU) and Beidou (China) constellations. In
particular, PAON4 sees a strong signal from Galileo E6 band, centred
at $1278.75\,\mathrm{MHz}$, as well as from Beidou in the same
band. Although GPS do not have direct emission in PAON4 band
($1250-1500\,\mathrm{MHz}$), we do see the GPS L2 signal centred
around $1227\,\mathrm{MHz}$, at the symmetric frequency of
$\sim1273\,\mathrm{MHz}$, with respect to the LO frequency at
$1250\,\mathrm{MHz}$. These satellite signals are used for determining
PAON4 geometry, beam shape, and instrumental phases.  Several satellite
tracks are usually present in our daily (24~hour) observations.  We
developed an additional software module included in our pipeline which
uses satellite tracks computed using the SGP4\footnote{\tt
  https://www.danrw.com/sgp4/} library from NORAD\footnote{\tt
  https://www.norad.mil} TLEs (Two Line Elements) available from
CELESTrak\footnote{\tt http://www.celestrak.com/NORAD/elements}, as
well as tracks of celestial objects at different frequencies,
separately for each polarization.  The module performs the following
operations:
\begin{enumerate}
\item Determine the pointing directions, as well as parametrised beam
  shape for each PAON4 antenna, using auto-correlation signals and the
  satellite tracks. It should be noted that the satellite signals are
  very strong.  Galileo satellites are a few hundred times brighter
  than Cas~A.
\item Using the parameters determined above, the phases are determined
  for each baseline separately. It is also possible to determine the
  phases from a combined fit over the six baselines and multiple
  frequencies using a linear frequency dependent phase model
  $\Phi_{ij}(\nu)$.
\item The derived instrumental phase variations as a function of pointing
  directions are used to refine PAON4 antenna positions along the vertical
  direction ($Oz$) and the North-South direction, as discussed
  in Section~\ref{Sec:Instrument-performance}.  The positions along the
  East-West direction changes the fringe rate and a combined fit,
  using visibility data from different declinations and many sources
  is used to obtain precise array geometry.
\item Among the fitted parameters are the source amplitudes.  The
  source amplitudes for Cas~A and Cyg A are used to determine PAON4
  radiometric calibration factors (Kelvin/NAU), to convert gain
  corrected visibilities to temperature. Instrument gain stability is
  also discussed in Section~\ref{Sec:Instrument-performance}.
\end{enumerate}

\section{Instrument performance }
\label{Sec:Instrument-performance} 
The analysis presented here is based on data taken in fall 2016, fall
2018, and winter and spring 2019.  The precision of calibration
parameters, stability, and overall instrument performance are
discussed here.

\subsection{System temperature and radiometric calibration stability}

The transit of Cas~A, Cyg~A, and satellites are used to determine the
antenna temperature of the 8 polarizations (H and V for each antenna).
Two procedures are used.  The first one, \textbf{(a)}, uses satellite
tracks and Cas~A and Cyg~A transits and are applied to data from the
2019 long duration $\mathrm{LD}$ observations. The dish diameter in
this procedure is determined from the autocorrelation signals using
satellite transits, while the bright source maximum amplitude $N_{ij}$
is obtained by fitting the cross-correlation signal between two
signals of the same polarization (ie. H-H or V-V cross-correlation
only). In the second procedure, \textbf{(b)}, the auto-correlation
signals during the Cas~A transit are used to derive the maximum signal
amplitude $N_i$ and the effective dish diameter $D_i$ for a given dish
and polarization. This second procedure is applied to the Cas~A
transits during the $\mathrm{Scan2018}$ (A,B) periods
(Table~\ref{Tab:listeobs}).  A Gaussian beam approximation is used in
the procedures.  Time-frequency maps produced after the gain
$g(\nu)\times G(t)$ correction lead to gain corrected autocorrelation
signal levels very close to 1~NAU, except during source or satellite
transits.


The fit coefficients are converted into radiometric calibration
coefficients $C^\mathrm{Jy}$ in units of $\mathrm{Jy/NAU}$ and
$C^\mathrm{K}$ in units of $\mathrm{Kelvin/NAU}$.  This is done using
the Cas~A and Cyg~A spectral flux reported in \cite{Perley_2017}.
Cas~A has an intensity of $I_s=1730$~Jy and $I_s=1772$~Jy at 1440~MHz
and 1396~MHz respectively.  Cyg~A has an intensity of $I_s=1585$~Jy at
1396~MHz.
The radiometric calibration coefficient
$C^\mathrm{K}$ in normalized units NAU is related to temperature in
Kelvin using the source intensity $I_s$ according to
\begin{equation}
C^\mathrm{K}_i = \frac{I_s}{2N_i}  \frac{\pi}{4k_B}\ D_i^2  \hspace{5mm} \mathrm{K/NAU} 
\end{equation}

\begin{table}
\caption{System temperature
  determined for each of the 8 polarizations using two
  different procedures \textbf{(a)} and \textbf{(b)} described in the text.}
\label{Tab:calib-tsys-casa}
\small
\centering
  \begin{tabular}{cccccc}
   \tableline
   Polar. &  $T_{a}^{CasA} (K)$ &  $T_{a}^{CygA} (K)$ &  $T_{a}^{CasA}(K)$  \\
             &  (a)                            &   (a)                          & (b)                 \\  \tableline
   1H   &   116                            &   147                        &144 \\ 
   2H   &   127                            &    129                       &133 \\ 
   3H   &   130                             &  122                        &138 \\ 
   4H   &   161                            &    152                       &185 \\ \tableline 
   1V   &   113                           &   100                         &   103\\ 
   2V   &    108                         &    92                           & 100 \\ 
   3V   &    103                          &   111                         & 98\\ 
   4V   &     143                         &  151                         & 180 \\ 
  \tableline
   \end{tabular}
\end{table}

With no strong sources in the field, signal levels are around 1~NAU.  The value of
the calibration coefficient is then equal to the system temperature
$\Tsys$ reported in Table~\ref{Tab:calib-tsys-casa}.  We have used two
procedures applied to different data sets to determine effective dish
diameters and radiometric calibration.

Procedure \textbf{(a)} operates at
1396~MHz {\color{black} to determine the system temperature. It uses the
  satellite tracks at 1278.75~MHz to fit the effective dish diameters
  and pointing parameters with the assumption that they do not vary
  too much over the bandwidth.}  The radiometric calibration is then
determined from fitting the 12~cross-correlation maximum intensities
during bright source transits.  This procedure is applied to
Cas~A and Cyg~A transits separately, during a single, although long
period of observation.

{\color{black} Procedure \textbf{(b)} is applied
  independently to visibilities at different frequencies, with results
  at 1440~MHz reported in Table~\ref{Tab:calib-tsys-casa}. It fits at
  once the maximum intensity, the effective diameter, and the azimuth
  misalignment of the dishes using the 8 auto-correlations during the
  8 Cas~A transits, with observations scattered over several
  months.  With method \textbf{(b)} we note that the effective diameter
  exhibits sinusoidal variations around $D\sim4.5\mathrm{m}$ as a
  function of frequency due to standing waves between the reflector
  and the feed.  These variations represent a $180^\circ$ phase shift
  between the H and V polarizations.}

Table~\ref{Tab:calib-tsys-casa} summarizes the different values of the
antenna temperatures.  We estimate a relative uncertainty of 10\%
mainly due to the maximum intensity and the effective diameter errors.
The stability level is better than 10\% based on measurements
performed over several months.

The results show a tendency for dish number 4 to be more noisy which
may be due to the fact that this dish is the closest to the
surrounding trees.  The H-polarization
feeds are about 20\% noisier than the V-polarization feeds ($H: \Tsys \sim
130 \; \mathrm{K}$, $V: \Tsys \sim 110 \; \mathrm{K}$). The origin of
this difference between H and V feeds is not yet understood.

The total noise temperature is $\sim 120 \; \mathrm{K}$.  Electronic
noise contributes $\sim 60-70 \; \mathrm{K}$ (see section
\ref{Sec:analogelectronic}).  Inefficiency of the mesh reflector adds
noise since it permits some ground thermal emission through.  Further
studies will be done to clarify the source of additional noise
contributions.

\subsection{Noise characterization}
Given the $\sim 10^5$ ratio between the noise level and the expected
cosmological 21~cm signal, which is essentially stochastic,
characterizing the noise and the way fluctuations decrease with
integration time is a central question for IM instruments.
Interferometry has the intrinsic advantage of being rather immune to
variations of individual receiver noise levels.  However, correlated
noise would limit the theoretical sensitivity expected from a
combination of visibility signals with long integration time.
Moreover, interferometric arrays designed for IM surveys are closely
packed, which increases the concern for the correlated noise between
receivers due to electromagnetic coupling between feeds.  We have
studied the noise behaviour of our complete electronic chain,
including the digitizer boards in laboratory, and identified some
coupling sources, responsible for correlated noise, such as the
coupling through the power supplies.

For more than a year up to July
2018, data were taken with a terminated 4V signal, where the input of
the amplifier on the 4V feed was connected to a $50\Omega$ resistor (see Section~\ref{Sec:Data-processing}).
These data were used to help identify the sources of various
perturbations, including the one subsequently identified as being due
to a bird.  They were also used to study
the correlated noise, comparing visibilities involving the 4V signal
with other signal pairs.

Our conclusions, from a preliminary study of
the correlated noise, are summarized below.

We analyzed data from the January 2018 observations and computed time
frequency maps with $\delta \nu \simeq 500 \, \mathrm{kHz}$ frequency
resolution and 12~minute time bins, corresponding to an effective per
pixel integration time of $\Delta t \sim 50 \mathrm{s}$.  Dispersions
on the visibility signals (real/imaginary parts) were then computed
over clean sections of the TFM maps, covering $\sim6$~hours in time
and $\sim35\,\mathrm{MHz}$, excluding bright regions of the sky.  A
noise reduction factor $\sqrt{\Delta t\times\delta\nu}\sim5000$ is
expected.  The r.m.s. $(\sigma)$ values quoted below assume that the
signals have been normalized to get auto-correlation levels equal to
unity (for fields without bright sources). 
\begin{itemize}
\item The noise level observed for signal pairs involving the
  terminated 4V signal (1V-4V , 1H-4V \ldots) are compatible with the
  levels expected given the integration time and frequency bandwidth,
  leading to $\sigma_{re,im}\sim2\times10^{-4}$.  The correlated
  noise contribution can be considered negligible, at least for the
  integration times of the current analysis.
\item For the same polarization signal pairs between two different
  dishes, such as (1H-3H) or (1V-3V), the observed noise level is
  higher compared to the expected level, assuming Gaussian
  uncorrelated noise. We measure dispersions up to 4-5 times higher,
  $\sigma_{re,im}\sim8\times10^{-4}$.
\item For two cross polarization probes located in the same feed/dish
  such as (1H-1V) or (3H-3V), the dispersion level increases to more
  than $ \sigma_{re,im}\sim2\times10^{-3}$.
\end{itemize}
The observed extra noise observed on the same polarization probes in
two different feeds, or two probes within the same feed shows up as
frequency dependent patterns, quite stable in time over a few
hours. In can be interpreted as noise generated in the analogue
electronic chain, cross fed through electromagnetic couplings between
feeds on two different dishes, or between probes in the same feed, and
then amplified in a way analogous to the Larsen effect.  Fortunately,
it can be efficiently subtracted due to its stability in time.
Further studies are needed to determine the noise floor due to this
correlated noise.

We have also analyzed how the noise level decreases with the
integration time.  Here, time frequency maps with different effective
per pixel integration times are computed, with a fixed frequency bin
width $\delta\nu=500\,\mathrm{kHz}$, from April-May 2019
data.  Maps with different averaging time window sizes, equal to
(1,2,4,8,16,32,64,128) in units of visibility sampling time are
built and then used to compute signal dispersions (in the absence of
bright source transits or RFI).  Data from spring 2019 was taken with a
visibility sampling time of 6~s leading to a maximum visibility
averaging time of 768~s.  PAON4 was
operated with $\sim10\%$ on sky efficiency so the maximum effective
integration time is slightly above one minute.

\begin{figure}
\centering
\hspace*{-0.1\columnwidth}
\begin{tabular}{cc}
\includegraphics[width=.8\columnwidth]{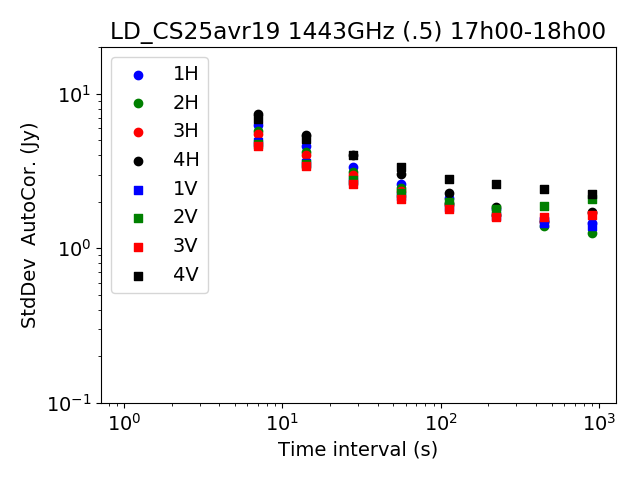}\\
\includegraphics[width=.8\columnwidth]{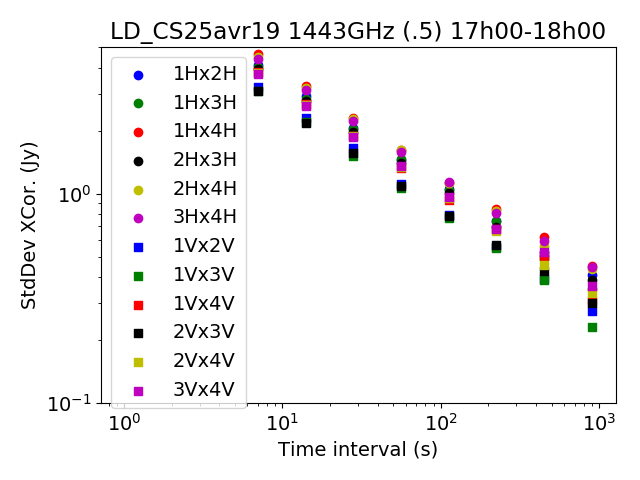}
\end{tabular}
\caption{Evolution of noise level with integration
  time. \textbf{Top:}~~Dispersion level for the 8 PAON4
  autocorrelation signals as a function of the integration time.
  \textbf{Bottom:}~~Dispersion level for the real parts of the 6 H-H
  and 6 V-V cross correlation signals as a function of integration
  time in seconds.}
\label{Fig:noisewtime}
\end{figure}

Figure \ref{Fig:noisewtime} shows the evolution of dispersion
levels computed as the r.m.s. of the different time-frequency maps,
excluding bright sources, satellites and RFI.  The correlated noise
is subtracted from the cross-correlation time-frequency maps,
using the frequency template obtained by the time averaged signal,
computed separately for each cross-correlation.  The left panel shows
the evolution of the dispersion with the integration time, for the 8
PAON4 autocorrelation signals.  The r.m.s. decreases with the
integration time, following a $1/\sqrt{\Delta t}$ trend. A noise
floor or saturation starts to appear at long integration times, more
or less strongly dependent on the data set and the auto-correlation
signal. However, it should be kept in mind that the r.m.s of the
autocorrelations signals is sensitive to gain variations
with time, as well to the variation of the diffuse sky brightness. The
presence of this apparent noise floor for the autocorrelation signals
does not imply that the underlying noise is not white.  The
right panel shows the evolution of the r.m.s. values computed on the
cross-correlation visibilities (real part) for the 12 H-H and V-V
correlations. Here, the r.m.s.  values are compatible with the
expected levels, and decrease with integration time following the
expected white noise law $1/\sqrt{\Delta t}$, without significant
contribution from correlated noise, once the average level at each
frequency has been subtracted.

\subsection{Phase calibration and array geometry}
\begin{figure}
\centering
\includegraphics[width=\columnwidth]{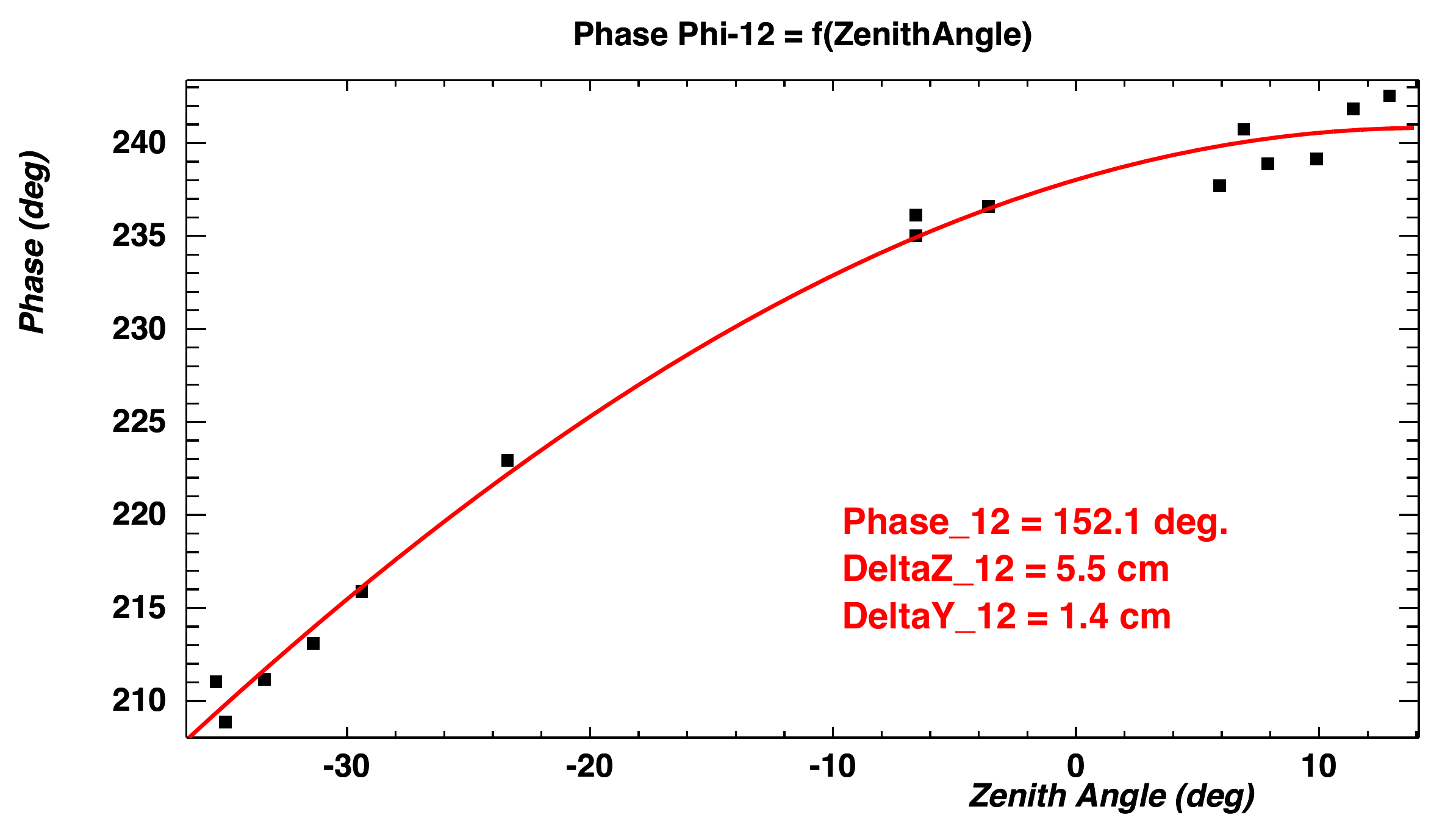}
\caption{Phase calibration and array geometry
  corrections. Instrumental phase values $\Phi_{12}$ determined for
  different scans, plotted as a function of the zenith angle (black
  squares). The red line represent a model fit, including corrections
  to the baseline and {\color{black} the residuals are within
    $\pm2^\circ$.}
}
\label{Fig:phasecorbasline}
\end{figure}

As mentioned in Section~\ref{Sec:on-sky-calibration}, Galileo
satellites are used to perform phase calibration and determine
instrumental phases for a large fraction of PAON4 observations carried
out in fall 2018, winter and spring 2019.  Instrumental phase values
are determined for each constant declination scan.
Figure~\ref{Fig:phasecorbasline} shows the antenna-2 phase
$\Delta\Phi_2=\Phi_{12}=\Phi_2-\Phi_1$ values, determined at
$1278.5\,\mathrm{MHz}$ for 17~scans and the H-polarization feeds, as a
function of the common, nominal antenna direction in the meridian
plane, referred to as the zenith angle
$(Z=90^\circ-\mathrm{elevation})$.  This angle corresponds to the
angle between the antenna axis and the local vertical, or the
difference between the observed declination and the instrument
latitude.  Negative zenith angles correspond to the antenna tilted
toward the south.  One can see that the phase $\Phi_{12}$ varies over
more than $30^\circ$, with a smooth variation as a function of the
zenith angle.  The variation is well explained by a shift in
theoretical antenna position, along the north-south $Oy$ and vertical
$Oz$ directions.  The fitted best model, taking into account the
baseline shift, is shown as the red curve. The fit result shows that
the two feed heights differ by about $\sim55\,\mathrm{mm}$ for this
(1H-2H) baseline, while the north-south component of this baseline
should be corrected by $\sim14\,\mathrm{mm}$.  Using the 6~baselines,
we determined the corrections to the array geometry, using the
zenith dependency of the instrumental phases.  We obtain a precision
of $\sim2\,\mathrm{mm}$ using these 17~scans. A shift in the
east-west baseline component would not show as a zenith angle phase
dependency, but rather as a change in the fringe rate.  We have not yet
determined baseline corrections along the east-west $Ox$
direction using the fringe rates, but higher precision is expected for
the determination of the east-west baseline corrections.
The instrumental phases do not change by more than $\pm2^\circ$
despite the fact that the observations were done over a period
spanning more than 8 months. Similar phase stability is observed for
the other baselines.

\subsection{Comparison of observed and expected signals}
\begin{figure*}
\centering
\begin{tabular}{cc}
\includegraphics[width=.5\textwidth]{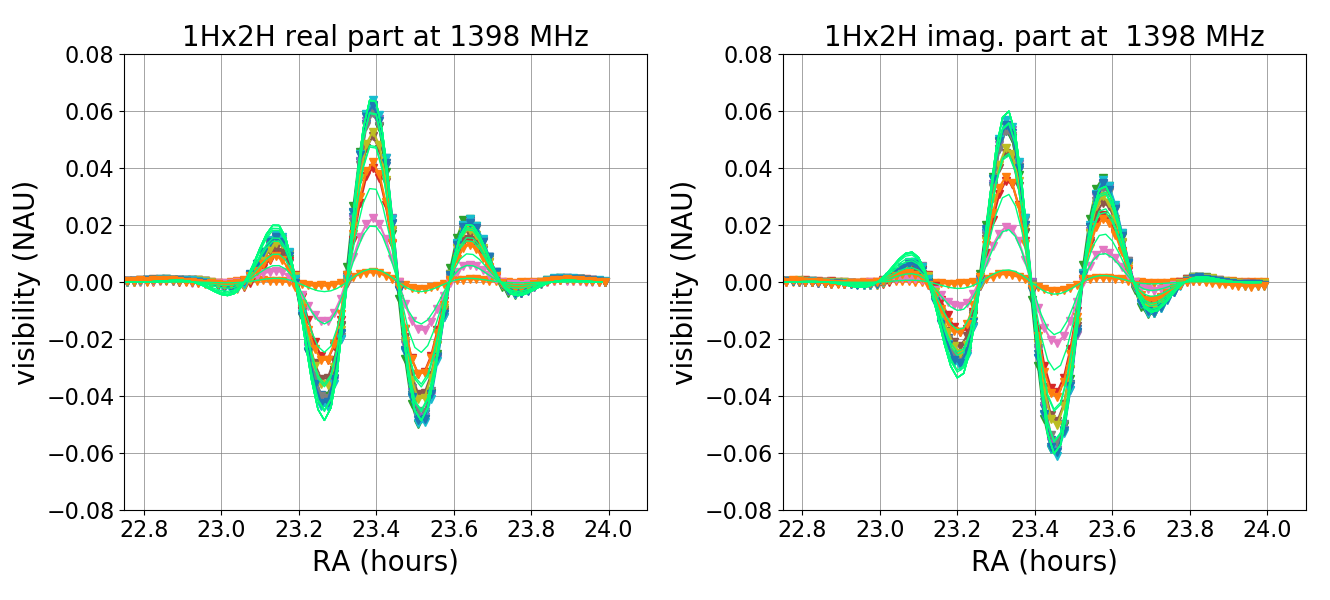}
&
\includegraphics[width=.5\textwidth]{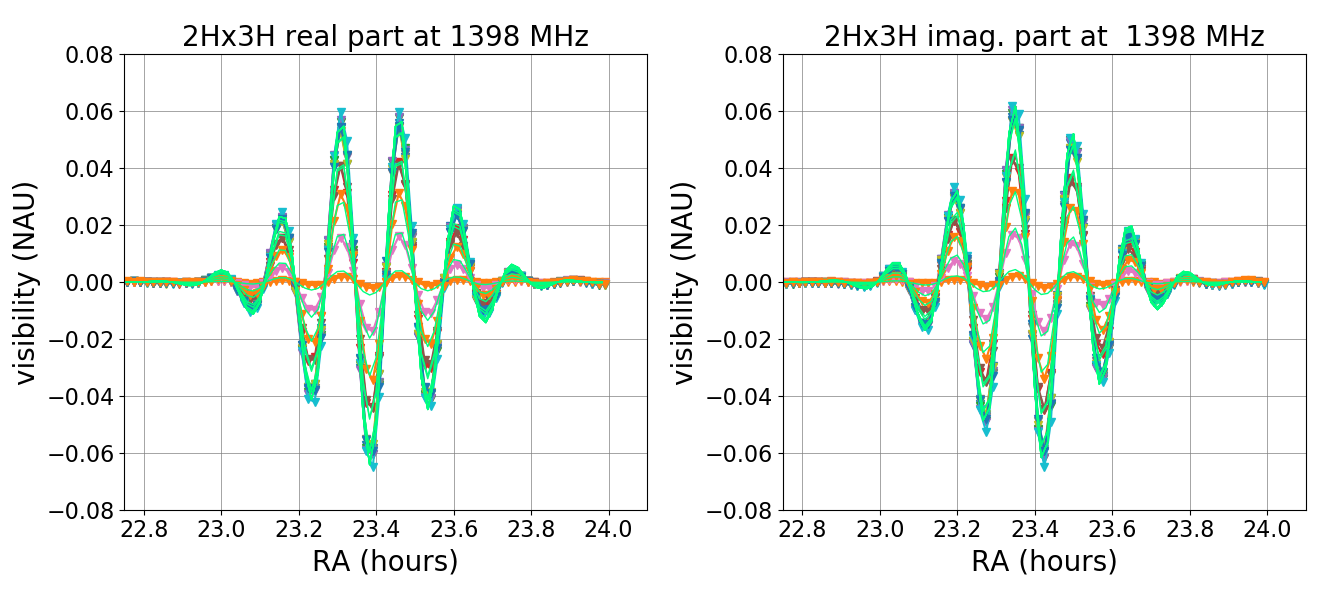} \\

\includegraphics[width=.5\textwidth]{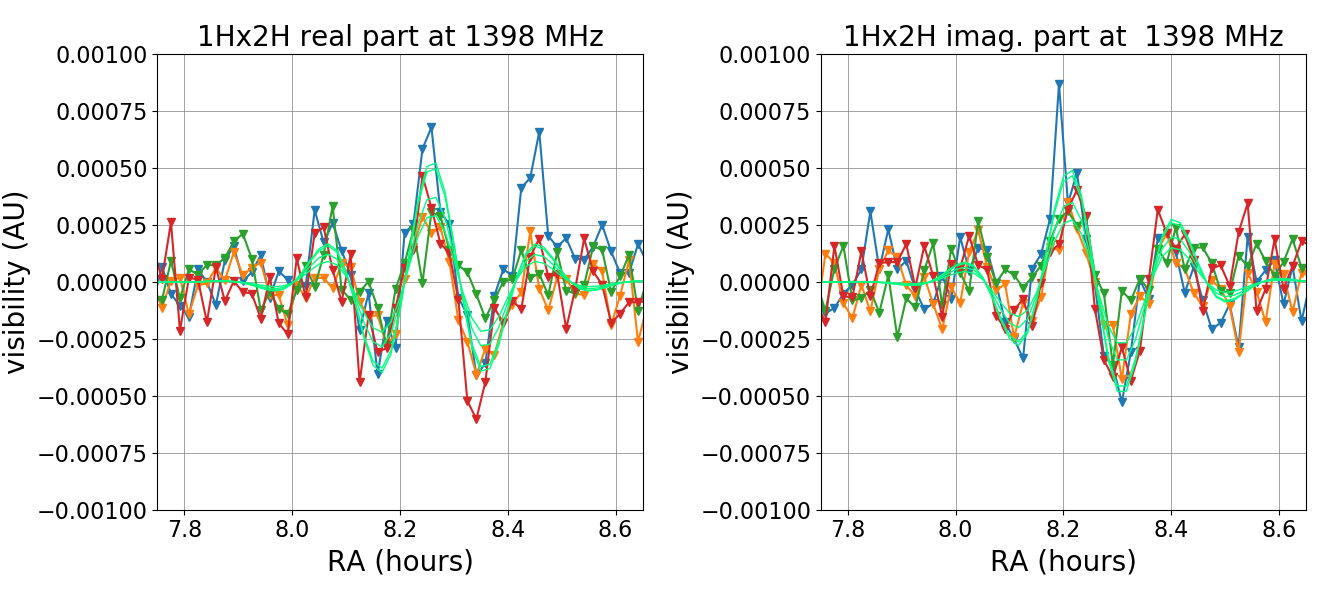}
&
\includegraphics[width=.5\textwidth]{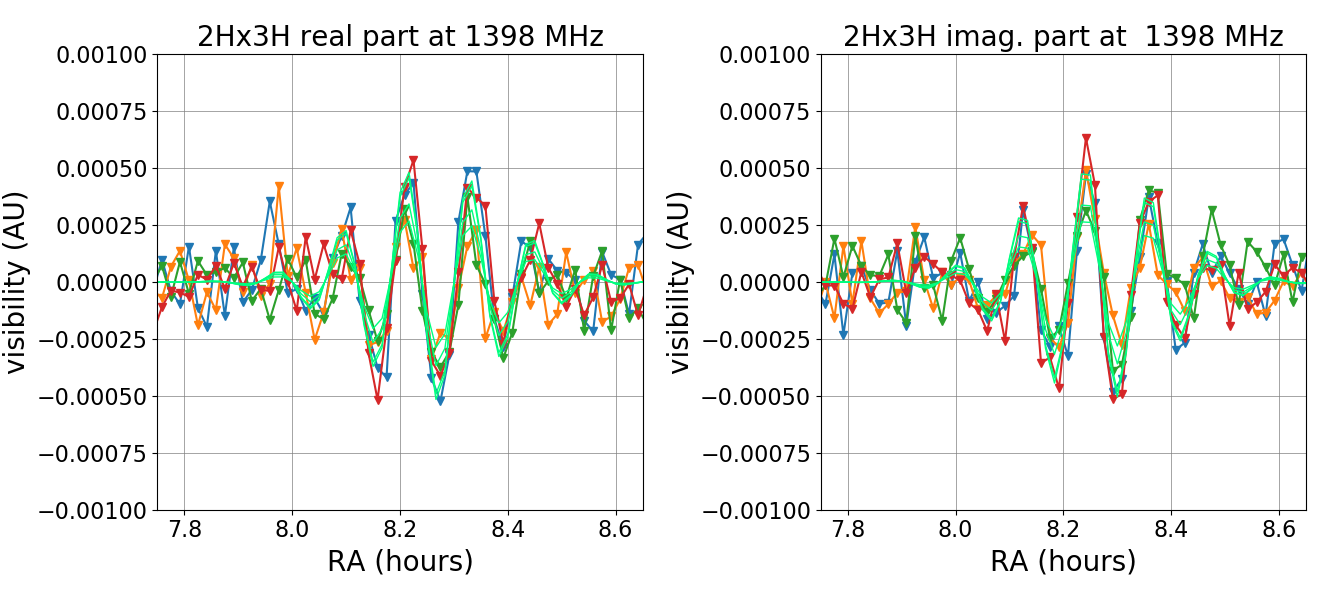} 
\end{tabular}

\caption{Comparisons of observed and expected (with a simplified
  model, see text) visibility variations with right ascension for
  scans near Cas~A (top) and 3C196 (bottom) declinations, from the
  Scan\_2018 A and B data, for the 1H-2H (left) and 2H-3H (right)
  cross-correlations. Each color corresponds to a different date, and
  the expected visibilities are shown in cyan.
\label{Fig:VisiSrcs}}
\end{figure*}

Figure~\ref{Fig:VisiSrcs} shows a comparison of observed and expected
visibilities for a few scans close to two bright sky sources, Cas~A
($\sim$1700~Jy) and 3C196 ($\sim$15~Jy) for the cross-correlation of
two pairs of feeds, 1H-2H and 2H-3H.  We gathered data from scans at
several declinations around the source, represented in different
colors. The visibility amplitude decreases for scans at more distant
declinations with respect to the source declination. This effect is
clearly visible Cas~A, but also for 3C196.  The expected signal was
rescaled using a single conversion coefficient per feed, used for both
sources, computed by adjusting the amplitudes of the expected Cas~A
on-source scan of July 17th, 2018.  This simplified computation of
expected signals does not take into account pointing uncertainties,
whereas a $\sim0.5\deg$ shift is suggested by our satellite fits.  Nor
does it take into account the non-Gaussian secondary lobes in the beam
pattern. The full signal level for Cas~A is expected to be about
$\sim6000\;\mathrm{mK}$, and $\sim60\;\mathrm{mK}$ for 3C196, while
the noise level is about $\sim20\;\mathrm{mK}$ given the time and
frequency binning used here.

\begin{figure} 
\centering
\hspace*{-3mm} \includegraphics[width=1.07\columnwidth]{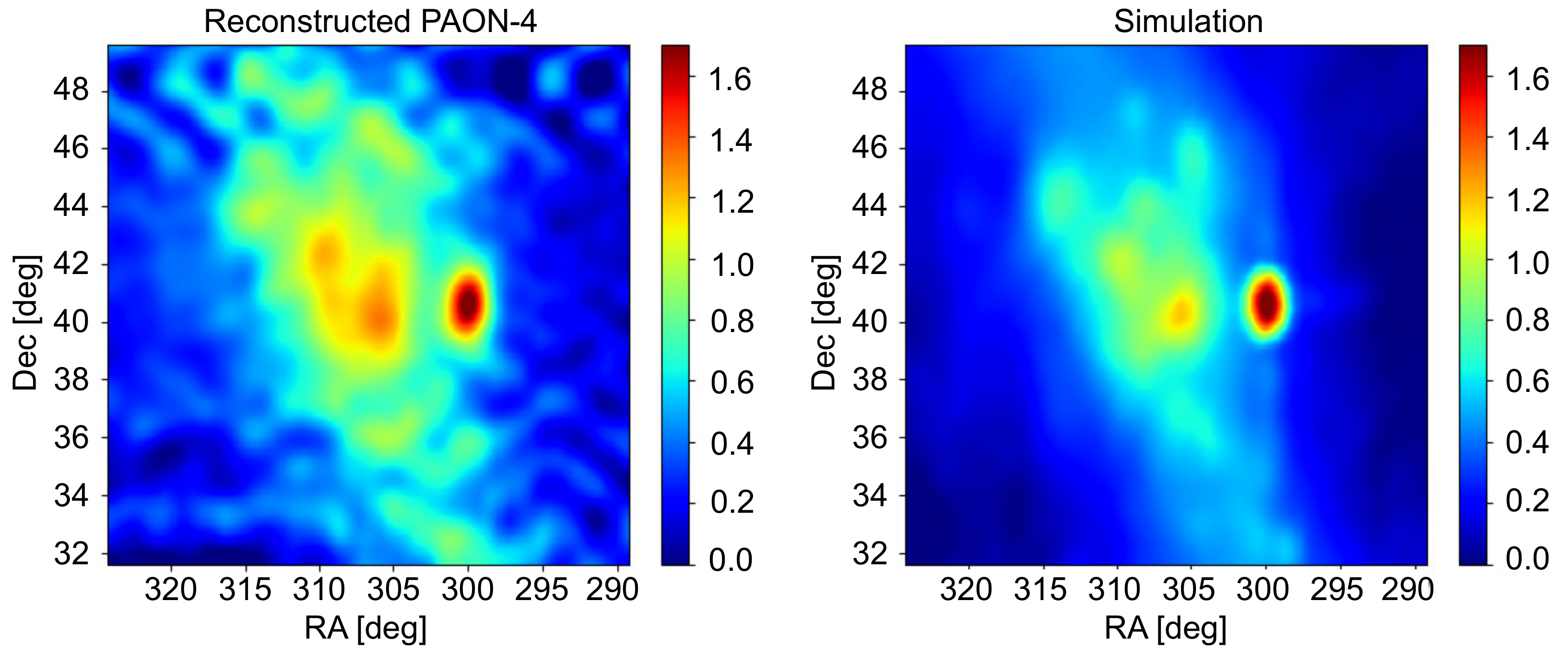} \\[1mm]
\caption{Example of a reconstructed map in a
  $\sim35^\circ\times18^\circ$ region around Cyg~A, covering the area
  $(32^\circ<\delta<50^\circ)$ in declination and
  $(290^\circ<\alpha<325^\circ)$ in right ascension, from November 2016
  data (left). Right panel shows the simulated map.\citep{2019PhDQHuang} }
\label{Fig:recmapCygAnov16}
\end{figure}

Figure~\ref{Fig:recmapCygAnov16} shows a region of the reconstructed
map at 1400~MHz using PAON4 observations from November 16th to
December, 1st 2016. 
A map making applying the
m-mode decomposition in harmonic space from transit visibility code, is used \citep{2019PhDQHuang}.  
A sky map covering the full 24~hour right
ascension, and the declination range
$35^\circ\lesssim\delta\lesssim46^\circ$ is computed from 11 24-hour
constant declination scans around the Cyg~A declination.  The
extracted map covers $\sim18^\circ$ in declination
$(32^\circ<\delta<50^\circ)$ and $\sim35^\circ$ in right ascension
$(290^\circ<\alpha<325^\circ)$ around the nearby radio galaxy
Cyg~A. The emission from Cyg~A, the Milky Way synchrotron emission, as
well as from the Cygnus~X star forming region, is clearly visible.

\section{Future prospects}
\label{Sec:Futur}


Time-variable systematics in the frequency response led us to develop
a new generation of sampling and signal processing board,
IDROGEN/NEBuLA, to perform digitisation as close to the feeds as
possible (see Section~\ref{Sec:spectral-response-time-gain}).  The
IDROGEN board is designed to equip interferometers with several
hundred feeds, scattered over a few hundred meters.  Clock
synchronisation is managed by the implementation of the White
Rabbit\footnote{\tt https://white-rabbit.web.cern.ch} technology.  A
first version of this new board, called NEBuLA (\textit{NumEriseur a
  Bande Large pour l'Astronomie}) was designed and produced in
2016-2017.  The second version, called IDROGEN, is developed as part
of the CNRS/IN2P3 DAQGEN project.  This project began in 2017 with the
goal of developing generic modules for rapid acquisition systems for
particle and astro-particle projects.

IDROGEN boards will be located in the electronic boxes on each
antenna, sampling the RF signals, relieving the need for frequency
shifting, and transmitting the digital streams over optical fibres all
the way to the computer cluster a few hundred meters away.  The
deployment of IDROGEN boards on PAON4 is foreseen in spring 2020,
after more in-depth characterisation of PAON4 in its present
configuration.  {\color{black} A reduction of the amplitude of the
  temperature dependent gain variations is also expected because the
  analogue chain is simplified and does not have the downconversion
  stage. It also has improved passive cooling.}

Upgrades of the acquisition system hardware and software are also
planned with the deployment of IDROGEN boards and should enable PAON4
to reach $ \sim 25 \%$ to $30\%$ on-sky time which is a significant
improvement compared to the current performance of $\lesssim 10\%$
(see Section~\ref{Sec:electronicandcorrelator}).  In particular, GPU
support will be added to the \textbf{TAcq} package for the correlation
computing software processor and also to the FFT processor to increase
the system throughput. First tests will be performed in the current
operation mode of PAON4.  At a later stage, the new digitiser boards
will be operated and qualified in the FFT mode.  After those tests, we
expect to release a stable version of the \textbf{TAcq} software, with
full support for IDROGEN boards.


\section{Conclusions}
\label{Sec:Conclusion}
A densely packed dish array interferometer is a cost effective option
to build radio instruments to survey large sky areas in L-band. We
have built and operated the PAON4 dish array transit interferometer
under very tight budget constraints.  Preliminary results indicate
that this type of instrument together with its associated observing
strategy are effective both in terms of scientific analysis and cost
effectiveness. The study presented here will progress further to
longer integration times in order to demonstrate that the expected
sensitivities at the few~mK level can be achieved.  In-depth
studies with more data and PAON4 maps will be published in the coming
year.  PAON4 does not have redundant baselines, and the additional
possibilities offered by the combination of nearly identical baselines
will be explored with larger instruments.   Such studies are being
pursued in parallel with the Tianlai dish array.  The next generation
IDROGEN digitizer/signal processor boards are in the final stage of
development and will be suited for a dish interferometer consisting of a few
hundred feeds and $\sim \mathrm{km}$ baselines.  IDROGEN will
be deployed for qualification on PAON4 in 2020 and we plan to carry a
higher sensitivity survey in the declination range $30^\circ \ldots
60^\circ$ in 2021.  {\color{black}The resulting 3D maps will be useful to
  characterize the spectral behaviour of the galactic foregrounds, including a determination of 
  its smoothness on the few~MHz scale.}

\section*{Acknowledgements}
The deployment of PAON4 and observations at Nan\c cay would not have
been possible without the help and support of the technical staff of
the Nan\c cay Radio Observatory, which is the Unit\'e scientifique de
Nan\c cay (USN) of the Observatoire de Paris, associated as Unit\'e de
Service et de Recherche (USR704) to the French Centre National de la
Recherche Scientifique (CNRS).  The Nan\c cay Observatory also
gratefully acknowledges the financial support of the Conseil
r\'egional of the R\' egion Centre in France. We acknowledge financial
support from ``Programme National de Cosmologie and Galaxies'' (PNCG)
of CNRS/INSU, France.



\newpage
\bibliographystyle{mnras}
\bibliography{Paon4p1}  

\begin{thebibliography}{}
\makeatletter
\relax
\def\mn@urlcharsother{\let\do\@makeother \do\$\do\&\do\#\do\^\do\_\do\%\do\~}
\def\mn@doi{\begingroup\mn@urlcharsother \@ifnextchar [ {\mn@doi@}
  {\mn@doi@[]}}
\def\mn@doi@[#1]#2{\def\@tempa{#1}\ifx\@tempa\@empty \href
  {http://dx.doi.org/#2} {doi:#2}\else \href {http://dx.doi.org/#2} {#1}\fi
  \endgroup}
\def\mn@eprint#1#2{\mn@eprint@#1:#2::\@nil}
\def\mn@eprint@arXiv#1{\href {http://arxiv.org/abs/#1} {{\tt arXiv:#1}}}
\def\mn@eprint@dblp#1{\href {http://dblp.uni-trier.de/rec/bibtex/#1.xml}
  {dblp:#1}}
\def\mn@eprint@#1:#2:#3:#4\@nil{\def\@tempa {#1}\def\@tempb {#2}\def\@tempc
  {#3}\ifx \@tempc \@empty \let \@tempc \@tempb \let \@tempb \@tempa \fi \ifx
  \@tempb \@empty \def\@tempb {arXiv}\fi \@ifundefined
  {mn@eprint@\@tempb}{\@tempb:\@tempc}{\expandafter \expandafter \csname
  mn@eprint@\@tempb\endcsname \expandafter{\@tempc}}}

\bibitem[\protect\citeauthoryear{{Abbott} et~al.,}{{Abbott}
  et~al.}{2018}]{2018MNRAS.tmp.3191A}
{Abbott} T.~M.~C.,  et~al., 2018, \mn@doi [\mnras] {10.1093/mnras/sty3351},
  \href {http://adsabs.harvard.edu/abs/2018MNRAS.tmp.3191A} {}

\bibitem[\protect\citeauthoryear{{Abdalla} \& {Rawlings}}{{Abdalla} \&
  {Rawlings}}{2005}]{2005MNRAS.360...27A}
{Abdalla} F.~B.,  {Rawlings} S.,  2005, \mn@doi [\mnras]
  {10.1111/j.1365-2966.2005.08650.x}, \href
  {http://adsabs.harvard.edu/abs/2005MNRAS.360...27A} {360, 27}

\bibitem[\protect\citeauthoryear{{Amendola}, {Kunz}, {Motta}, {Saltas}  \&
  {Sawicki}}{{Amendola} et~al.}{2013}]{2013PhRvD..87b3501A}
{Amendola} L.,  {Kunz} M.,  {Motta} M.,  {Saltas} I.~D.,   {Sawicki} I.,  2013,
  \mn@doi [\prd] {10.1103/PhysRevD.87.023501}, \href
  {http://adsabs.harvard.edu/abs/2013PhRvD..87b3501A} {87, 023501}

\bibitem[\protect\citeauthoryear{{Ansari}, {Le Goff}, {Magneville}, {Moniez},
  {Palanque-Delabrouille}, {Rich}, {Ruhlmann-Kleider}  \& {Y{\`e}che}}{{Ansari}
  et~al.}{2008}]{2008arXiv0807.3614A}
{Ansari} R.,  {Le Goff} J.~.,  {Magneville} C.,  {Moniez} M.,
  {Palanque-Delabrouille} N.,  {Rich} J.,  {Ruhlmann-Kleider} V.,   {Y{\`e}che}
  C.,  2008, preprint, \href
  {http://adsabs.harvard.edu/abs/2008arXiv0807.3614A} {} (\mn@eprint {arXiv}
  {0807.3614})

\bibitem[\protect\citeauthoryear{{Ansari}, {Campagne}, {Colom}, {Magneville},
  {Martin}, {Moniez}, {Rich}  \& {Y{\`e}che}}{{Ansari}
  et~al.}{2012a}]{2012CRPhy..13...46A}
{Ansari} R.,  {Campagne} J.-E.,  {Colom} P.,  {Magneville} C.,  {Martin} J.-M.,
   {Moniez} M.,  {Rich} J.,   {Y{\`e}che} C.,  2012a, \mn@doi [Comptes Rendus
  Physique] {10.1016/j.crhy.2011.11.003}, \href
  {http://adsabs.harvard.edu/abs/2012CRPhy..13...46A} {13, 46}

\bibitem[\protect\citeauthoryear{{Ansari} et~al.,}{{Ansari}
  et~al.}{2012b}]{A&A...540A.129A}
{Ansari} R.,  et~al., 2012b, \mn@doi [\aap] {10.1051/0004-6361/201117837},
  \href {http://adsabs.harvard.edu/abs/2012A%26A...540A.129A} {540, A129}

\bibitem[\protect\citeauthoryear{{Ansari}, {Campagne}, {Colom}, {Ferrari},
  {Magneville}, {Martin}, {Moniez}  \& {Torrent{\'o}}}{{Ansari}
  et~al.}{2016}]{2016ExA....41..117A}
{Ansari} R.,  {Campagne} J.~E.,  {Colom} P.,  {Ferrari} C.,  {Magneville} C.,
  {Martin} J.~M.,  {Moniez} M.,   {Torrent{\'o}} A.~S.,  2016, \mn@doi
  [Experimental Astronomy] {10.1007/s10686-015-9477-7}, \href
  {http://adsabs.harvard.edu/abs/2016ExA....41..117A} {41, 117}

\bibitem[\protect\citeauthoryear{{Armstrong}, {Hickish}, {Zarb Adami}  \&
  {Jones}}{{Armstrong} et~al.}{2009}]{2009wska.confE..45A}
{Armstrong} R.,  {Hickish} J.,  {Zarb Adami} K.,   {Jones} M.~E.,  2009, in
  Wide Field Astronomy \&amp; Technology for the Square Kilometre Array. p.~45
  (\mn@eprint {arXiv} {0912.0380}), \mn@doi{10.22323/1.132.0045}

\bibitem[\protect\citeauthoryear{{Bandura}}{{Bandura}}{2011}]{2011PhDT.......158B}
{Bandura} K.,  2011, PhD thesis, Carnegie Mellon University

\bibitem[\protect\citeauthoryear{{Bandura} et~al.,}{{Bandura}
  et~al.}{2014}]{2014SPIE.9145E..22B}
{Bandura} K.,  et~al., 2014, in Ground-based and Airborne Telescopes V. p.
  914522 (\mn@eprint {arXiv} {1406.2288}), \mn@doi{10.1117/12.2054950}

\bibitem[\protect\citeauthoryear{{Bandura} et~al.,}{{Bandura}
  et~al.}{2019}]{2019arXiv190712559B}
{Bandura} K.,  et~al., 2019, arXiv e-prints, \href
  {https://ui.adsabs.harvard.edu/abs/2019arXiv190712559B} {p. arXiv:1907.12559}

\bibitem[\protect\citeauthoryear{{Battye}, {Browne}, {Dickinson}, {Heron},
  {Maffei}  \& {Pourtsidou}}{{Battye} et~al.}{2013}]{2013MNRAS.434.1239B}
{Battye} R.~A.,  {Browne} I.~W.~A.,  {Dickinson} C.,  {Heron} G.,  {Maffei} B.,
    {Pourtsidou} A.,  2013, \mn@doi [\mnras] {10.1093/mnras/stt1082}, \href
  {http://adsabs.harvard.edu/abs/2013MNRAS.434.1239B} {434, 1239}

\bibitem[\protect\citeauthoryear{Betoule et~al.}{Betoule
  et~al.}{2014}]{Betoule:2014frx}
Betoule M.,  et~al., 2014, \mn@doi [Astron. Astrophys.]
  {10.1051/0004-6361/201423413}, 568, A22

\bibitem[\protect\citeauthoryear{{Bigot-Sazy} et~al.,}{{Bigot-Sazy}
  et~al.}{2015}]{2015MNRAS.454.3240B}
{Bigot-Sazy} M.~A.,  et~al., 2015, \mn@doi [\mnras] {10.1093/mnras/stv2153},
  \href {https://ui.adsabs.harvard.edu/abs/2015MNRAS.454.3240B} {454, 3240}

\bibitem[\protect\citeauthoryear{{Bigot-Sazy} et~al.,}{{Bigot-Sazy}
  et~al.}{2016}]{2016ASPC..502...41B}
{Bigot-Sazy} M.-A.,  et~al., 2016, in {Qain} L.,  {Li} D.,  eds,  Astronomical
  Society of the Pacific Conference Series Vol. 502, Frontiers in Radio
  Astronomy and FAST Early Sciences Symposium 2015. p.~41 (\mn@eprint {arXiv}
  {1511.03006})

\bibitem[\protect\citeauthoryear{{Braun}, {Bourke}, {Green}, {Keane}  \&
  {Wagg}}{{Braun} et~al.}{2015}]{2015aska.confE.174B}
{Braun} R.,  {Bourke} T.,  {Green} J.~A.,  {Keane} E.,   {Wagg} J.,  2015,
  Advancing Astrophysics with the Square Kilometre Array (AASKA14), \href
  {http://adsabs.harvard.edu/abs/2015aska.confE.174B} {p.~174}

\bibitem[\protect\citeauthoryear{{Bull}, {Camera}, {Raccanelli}, {Blake},
  {Ferreira}, {Santos}  \& {Schwarz}}{{Bull}
  et~al.}{2015a}]{2015aska.confE..24B}
{Bull} P.,  {Camera} S.,  {Raccanelli} A.,  {Blake} C.,  {Ferreira} P.,
  {Santos} M.,   {Schwarz} D.~J.,  2015a, Advancing Astrophysics with the
  Square Kilometre Array (AASKA14), \href
  {http://adsabs.harvard.edu/abs/2015aska.confE..24B} {p.~24}

\bibitem[\protect\citeauthoryear{{Bull}, {Ferreira}, {Patel}  \&
  {Santos}}{{Bull} et~al.}{2015b}]{2015ApJ...803...21B}
{Bull} P.,  {Ferreira} P.~G.,  {Patel} P.,   {Santos} M.~G.,  2015b, \mn@doi
  [\apj] {10.1088/0004-637X/803/1/21}, \href
  {http://adsabs.harvard.edu/abs/2015ApJ...803...21B} {803, 21}

\bibitem[\protect\citeauthoryear{{Bull} et~al.,}{{Bull}
  et~al.}{2016}]{2016PDU....12...56B}
{Bull} P.,  et~al., 2016, \mn@doi [Physics of the Dark Universe]
  {10.1016/j.dark.2016.02.001}, \href
  {http://adsabs.harvard.edu/abs/2016PDU....12...56B} {12, 56}

\bibitem[\protect\citeauthoryear{{Byrne} et~al.,}{{Byrne}
  et~al.}{2019}]{2019ApJ...875...70B}
{Byrne} R.,  et~al., 2019, \mn@doi [\apj] {10.3847/1538-4357/ab107d}, \href
  {https://ui.adsabs.harvard.edu/abs/2019ApJ...875...70B} {875, 70}

\bibitem[\protect\citeauthoryear{{Chang} \& {GBT-HIM Team}}{{Chang} \& {GBT-HIM
  Team}}{2016}]{2016AAS...22742601C}
{Chang} T.-C.,  {GBT-HIM Team} 2016, in American Astronomical Society Meeting
  Abstracts \#227. p. 426.01

\bibitem[\protect\citeauthoryear{{Chang}, {Pen}, {Peterson}  \&
  {McDonald}}{{Chang} et~al.}{2008}]{2008PhRvL.100i1303C}
{Chang} T.-C.,  {Pen} U.-L.,  {Peterson} J.~B.,   {McDonald} P.,  2008, \mn@doi
  [Physical Review Letters] {10.1103/PhysRevLett.100.091303}, \href
  {http://adsabs.harvard.edu/abs/2008PhRvL.100i1303C} {100, 091303}

\bibitem[\protect\citeauthoryear{{Chang}, {Pen}, {Bandura}  \&
  {Peterson}}{{Chang} et~al.}{2010}]{2010Natur.466..463C}
{Chang} T.-C.,  {Pen} U.-L.,  {Bandura} K.,   {Peterson} J.~B.,  2010, \mn@doi
  [\nat] {10.1038/nature09187}, \href
  {http://adsabs.harvard.edu/abs/2010Natur.466..463C} {466, 463}

\bibitem[\protect\citeauthoryear{{Charlet} et~al.,}{{Charlet}
  et~al.}{2011}]{2011ITNS...58.1833C}
{Charlet} D.,  et~al., 2011, \mn@doi [IEEE Transactions on Nuclear Science]
  {10.1109/TNS.2011.2155085}, \href
  {http://adsabs.harvard.edu/abs/2011ITNS...58.1833C} {58, 1833}

\bibitem[\protect\citeauthoryear{{Chatterjee} et~al.,}{{Chatterjee}
  et~al.}{2017}]{2017Natur.541...58C}
{Chatterjee} S.,  et~al., 2017, \mn@doi [\nat] {10.1038/nature20797}, \href
  {https://ui.adsabs.harvard.edu/abs/2017Natur.541...58C} {541, 58}

\bibitem[\protect\citeauthoryear{{Chen}}{{Chen}}{2012}]{2012IJMPS..12..256C}
{Chen} X.,  2012, in International Journal of Modern Physics Conference Series.
  pp 256--263 (\mn@eprint {arXiv} {1212.6278}),
  \mn@doi{10.1142/S2010194512006459}

\bibitem[\protect\citeauthoryear{{Cosmic Visions 21 cm Collaboration}
  et~al.,}{{Cosmic Visions 21 cm Collaboration}
  et~al.}{2018}]{2018arXiv181009572C}
{Cosmic Visions 21 cm Collaboration} et~al., 2018, arXiv e-prints, \href
  {http://adsabs.harvard.edu/abs/2018arXiv181009572C} {}

\bibitem[\protect\citeauthoryear{{DES Collaboration} et~al.,}{{DES
  Collaboration} et~al.}{2018}]{2018arXiv181102375D}
{DES Collaboration} et~al., 2018, arXiv e-prints, \href
  {http://adsabs.harvard.edu/abs/2018arXiv181102375D} {}

\bibitem[\protect\citeauthoryear{{Das} et~al.,}{{Das}
  et~al.}{2018}]{2018SPIE10708E..36D}
{Das} S.,  et~al., 2018, in Millimeter, Submillimeter, and Far-Infrared
  Detectors and Instrumentation for Astronomy IX. p. 1070836 (\mn@eprint
  {arXiv} {1806.04698}), \mn@doi{10.1117/12.2313031}

\bibitem[\protect\citeauthoryear{{Deschamps} et~al.,}{{Deschamps}
  et~al.}{2013}]{6588971}
{Deschamps} H.,  et~al., 2013, \mn@doi [IEEE Transactions on Nuclear Science]
  {10.1109/TNS.2013.2277663}, 60, 3620

\bibitem[\protect\citeauthoryear{{Furlanetto}, {Oh}  \& {Briggs}}{{Furlanetto}
  et~al.}{2006}]{2006PhR...433..181F}
{Furlanetto} S.~R.,  {Oh} S.~P.,   {Briggs} F.~H.,  2006, \mn@doi [\physrep]
  {10.1016/j.physrep.2006.08.002}, \href
  {http://adsabs.harvard.edu/abs/2006PhR...433..181F} {433, 181}

\bibitem[\protect\citeauthoryear{{Hamaker}, {Bregman}  \& {Sault}}{{Hamaker}
  et~al.}{1996}]{1996A&AS..117..137H}
{Hamaker} J.~P.,  {Bregman} J.~D.,   {Sault} R.~J.,  1996, \aaps, \href
  {https://ui.adsabs.harvard.edu/abs/1996A%26AS..117..137H} {117, 137}

\bibitem[\protect\citeauthoryear{{Hinshaw} et~al.,}{{Hinshaw}
  et~al.}{2013}]{2013ApJS..208...19H}
{Hinshaw} G.,  et~al., 2013, \mn@doi [\apjs] {10.1088/0067-0049/208/2/19},
  \href {http://adsabs.harvard.edu/abs/2013ApJS..208...19H} {208, 19}

\bibitem[\protect\citeauthoryear{{Huang}}{{Huang}}{2019}]{2019PhDQHuang}
{Huang} Q.,  2019, PhD thesis, Paris-Saclay University \& University of Chinese
  Academy of Science

\bibitem[\protect\citeauthoryear{{Kazemi}, {Yatawatta}, {Zaroubi},
  {Lampropoulos}, {de Bruyn}, {Koopmans}  \& {Noordam}}{{Kazemi}
  et~al.}{2011}]{2011MNRAS.414.1656K}
{Kazemi} S.,  {Yatawatta} S.,  {Zaroubi} S.,  {Lampropoulos} P.,  {de Bruyn}
  A.~G.,  {Koopmans} L.~V.~E.,   {Noordam} J.,  2011, \mn@doi [\mnras]
  {10.1111/j.1365-2966.2011.18506.x}, \href
  {https://ui.adsabs.harvard.edu/abs/2011MNRAS.414.1656K} {414, 1656}

\bibitem[\protect\citeauthoryear{{Masui} et~al.,}{{Masui}
  et~al.}{2013}]{2013ApJ...763L..20M}
{Masui} K.~W.,  et~al., 2013, \mn@doi [\apjl] {10.1088/2041-8205/763/1/L20},
  \href {http://adsabs.harvard.edu/abs/2013ApJ...763L..20M} {763, L20}

\bibitem[\protect\citeauthoryear{{Newburgh} et~al.,}{{Newburgh}
  et~al.}{2016}]{2016SPIE.9906E..5XN}
{Newburgh} L.~B.,  et~al., 2016, in Ground-based and Airborne Telescopes VI. p.
  99065X (\mn@eprint {arXiv} {1607.02059}), \mn@doi{10.1117/12.2234286}

\bibitem[\protect\citeauthoryear{{Parsons} et~al.,}{{Parsons}
  et~al.}{2010}]{2010AJ....139.1468P}
{Parsons} A.~R.,  et~al., 2010, \mn@doi [\aj] {10.1088/0004-6256/139/4/1468},
  \href {http://adsabs.harvard.edu/abs/2010AJ....139.1468P} {139, 1468}

\bibitem[\protect\citeauthoryear{{Perera} et~al.,}{{Perera}
  et~al.}{2019}]{2019MNRAS.490.4666P}
{Perera} B.~B.~P.,  et~al., 2019, \mn@doi [\mnras] {10.1093/mnras/stz2857},
  \href {https://ui.adsabs.harvard.edu/abs/2019MNRAS.490.4666P} {490, 4666}

\bibitem[\protect\citeauthoryear{Perley \& Butler}{Perley \&
  Butler}{2017}]{Perley_2017}
Perley R.~A.,  Butler B.~J.,  2017, \mn@doi [The Astrophysical Journal
  Supplement Series] {10.3847/1538-4365/aa6df9}, 230, 7

\bibitem[\protect\citeauthoryear{{Peterson}, {Bandura}  \& {Pen}}{{Peterson}
  et~al.}{2006}]{2006astro.ph..6104P}
{Peterson} J.~B.,  {Bandura} K.,   {Pen} U.~L.,  2006, arXiv Astrophysics
  e-prints, \href {http://adsabs.harvard.edu/abs/2006astro.ph..6104P} {}

\bibitem[\protect\citeauthoryear{{Peterson} et~al.,}{{Peterson}
  et~al.}{2009}]{2009astro2010S.234P}
{Peterson} J.~B.,  et~al., 2009, in astro2010: The Astronomy and Astrophysics
  Decadal Survey.  (\mn@eprint {arXiv} {0902.3091})

\bibitem[\protect\citeauthoryear{{Planck Collaboration} et~al.,}{{Planck
  Collaboration} et~al.}{2016}]{2016A&A...594A..13P}
{Planck Collaboration} et~al., 2016, \mn@doi [\aap]
  {10.1051/0004-6361/201525830}, \href
  {http://adsabs.harvard.edu/abs/2016A%26A...594A..13P} {594, A13}

\bibitem[\protect\citeauthoryear{{Planck Collaboration} et~al.,}{{Planck
  Collaboration} et~al.}{2018}]{2018arXiv180706209P}
{Planck Collaboration} et~al., 2018, arXiv e-prints, \href
  {https://ui.adsabs.harvard.edu/abs/2018arXiv180706209P} {p. arXiv:1807.06209}

\bibitem[\protect\citeauthoryear{{Pober} et~al.,}{{Pober}
  et~al.}{2014}]{2014ApJ...782...66P}
{Pober} J.~C.,  et~al., 2014, \mn@doi [\apj] {10.1088/0004-637X/782/2/66},
  \href {http://adsabs.harvard.edu/abs/2014ApJ...782...66P} {782, 66}

\bibitem[\protect\citeauthoryear{{Pritchard} \& {Loeb}}{{Pritchard} \&
  {Loeb}}{2008}]{2008PhRvD..78j3511P}
{Pritchard} J.~R.,  {Loeb} A.,  2008, \mn@doi [\prd]
  {10.1103/PhysRevD.78.103511}, \href
  {http://adsabs.harvard.edu/abs/2008PhRvD..78j3511P} {78, 103511}

\bibitem[\protect\citeauthoryear{{Salazar-Albornoz} et~al.,}{{Salazar-Albornoz}
  et~al.}{2017}]{2017MNRAS.468.2938S}
{Salazar-Albornoz} S.,  et~al., 2017, \mn@doi [\mnras] {10.1093/mnras/stx633},
  \href {http://adsabs.harvard.edu/abs/2017MNRAS.468.2938S} {468, 2938}

\bibitem[\protect\citeauthoryear{{Salvini} \& {Wijnholds}}{{Salvini} \&
  {Wijnholds}}{2014}]{2014A&A...571A..97S}
{Salvini} S.,  {Wijnholds} S.~J.,  2014, \mn@doi [\aap]
  {10.1051/0004-6361/201424487}, \href
  {https://ui.adsabs.harvard.edu/abs/2014A&A...571A..97S} {571, A97}

\bibitem[\protect\citeauthoryear{{Sault}, {Hamaker}  \& {Bregman}}{{Sault}
  et~al.}{1996}]{1996A&AS..117..149S}
{Sault} R.~J.,  {Hamaker} J.~P.,   {Bregman} J.~D.,  1996, \aaps, \href
  {https://ui.adsabs.harvard.edu/abs/1996A%26AS..117..149S} {117, 149}

\bibitem[\protect\citeauthoryear{{Shaw}, {Sigurdson}, {Pen}, {Stebbins}  \&
  {Sitwell}}{{Shaw} et~al.}{2014}]{2014ApJ...781...57S}
{Shaw} J.~R.,  {Sigurdson} K.,  {Pen} U.-L.,  {Stebbins} A.,   {Sitwell} M.,
  2014, \mn@doi [\apj] {10.1088/0004-637X/781/2/57}, \href
  {http://adsabs.harvard.edu/abs/2014ApJ...781...57S} {781, 57}

\bibitem[\protect\citeauthoryear{{Shaw}, {Sigurdson}, {Sitwell}, {Stebbins}  \&
  {Pen}}{{Shaw} et~al.}{2015}]{2015PhRvD..91h3514S}
{Shaw} J.~R.,  {Sigurdson} K.,  {Sitwell} M.,  {Stebbins} A.,   {Pen} U.-L.,
  2015, \mn@doi [\prd] {10.1103/PhysRevD.91.083514}, \href
  {http://adsabs.harvard.edu/abs/2015PhRvD..91h3514S} {91, 083514}

\bibitem[\protect\citeauthoryear{{Smirnov}}{{Smirnov}}{2011}]{2011A&A...527A.106S}
{Smirnov} O.~M.,  2011, \mn@doi [\aap] {10.1051/0004-6361/201016082}, \href
  {https://ui.adsabs.harvard.edu/abs/2011A&A...527A.106S} {527, A106}

\bibitem[\protect\citeauthoryear{Smirnov \& Tasse}{Smirnov \&
  Tasse}{2015}]{10.1093/mnras/stv418}
Smirnov O.~M.,  Tasse C.,  2015, \mn@doi [Monthly Notices of the Royal
  Astronomical Society] {10.1093/mnras/stv418}, 449, 2668

\bibitem[\protect\citeauthoryear{{Stovall} et~al.,}{{Stovall}
  et~al.}{2014}]{2014ApJ...791...67S}
{Stovall} K.,  et~al., 2014, \mn@doi [\apj] {10.1088/0004-637X/791/1/67}, \href
  {https://ui.adsabs.harvard.edu/abs/2014ApJ...791...67S} {791, 67}

\bibitem[\protect\citeauthoryear{{Tegmark} \& {Zaldarriaga}}{{Tegmark} \&
  {Zaldarriaga}}{2009}]{2009PhRvD..79h3530T}
{Tegmark} M.,  {Zaldarriaga} M.,  2009, \mn@doi [\prd]
  {10.1103/PhysRevD.79.083530}, \href
  {http://adsabs.harvard.edu/abs/2009PhRvD..79h3530T} {79, 083530}

\bibitem[\protect\citeauthoryear{{Thornton} et~al.,}{{Thornton}
  et~al.}{2013}]{2013Sci...341...53T}
{Thornton} D.,  et~al., 2013, \mn@doi [Science] {10.1126/science.1236789},
  \href {https://ui.adsabs.harvard.edu/abs/2013Sci...341...53T} {341, 53}

\bibitem[\protect\citeauthoryear{{Tingay} et~al.,}{{Tingay}
  et~al.}{2013}]{2013PASA...30....7T}
{Tingay} S.~J.,  et~al., 2013, \mn@doi [\pasa] {10.1017/pasa.2012.007}, \href
  {http://adsabs.harvard.edu/abs/2013PASA...30....7T} {30, e007}

\bibitem[\protect\citeauthoryear{{Torchinsky}, {Olofsson}, {Censier},
  {Karastergiou}, {Serylak}, {Picard}, {Renaud}  \& {Taffoureau}}{{Torchinsky}
  et~al.}{2016}]{2016A&A...589A..77T}
{Torchinsky} S.~A.,  {Olofsson} A.~O.~H.,  {Censier} B.,  {Karastergiou} A.,
  {Serylak} M.,  {Picard} P.,  {Renaud} P.,   {Taffoureau} C.,  2016, \mn@doi
  [\aap] {10.1051/0004-6361/201526706}, \href
  {http://adsabs.harvard.edu/abs/2016A%26A...589A..77T} {589, A77}

\bibitem[\protect\citeauthoryear{{Wang}, {Tegmark}, {Santos}  \& {Knox}}{{Wang}
  et~al.}{2006}]{2006ApJ...650..529W}
{Wang} X.,  {Tegmark} M.,  {Santos} M.~G.,   {Knox} L.,  2006, \mn@doi [\apj]
  {10.1086/506597}, \href {http://adsabs.harvard.edu/abs/2006ApJ...650..529W}
  {650, 529}

\bibitem[\protect\citeauthoryear{{Wolz}, {Abdalla}, {Blake}, {Shaw}, {Chapman}
  \& {Rawlings}}{{Wolz} et~al.}{2014}]{2014MNRAS.441.3271W}
{Wolz} L.,  {Abdalla} F.~B.,  {Blake} C.,  {Shaw} J.~R.,  {Chapman} E.,
  {Rawlings} S.,  2014, \mn@doi [\mnras] {10.1093/mnras/stu792}, \href
  {http://adsabs.harvard.edu/abs/2014MNRAS.441.3271W} {441, 3271}

\bibitem[\protect\citeauthoryear{{Wuensche} \& {the BINGO
  Collaboration}}{{Wuensche} \& {the BINGO
  Collaboration}}{2018}]{2018arXiv180301644W}
{Wuensche} C.~A.,  {the BINGO Collaboration} 2018, arXiv e-prints, \href
  {https://ui.adsabs.harvard.edu/abs/2018arXiv180301644W} {p. arXiv:1803.01644}

\bibitem[\protect\citeauthoryear{{Wyithe}, {Loeb}  \& {Geil}}{{Wyithe}
  et~al.}{2008}]{2008MNRAS.383.1195W}
{Wyithe} J.~S.~B.,  {Loeb} A.,   {Geil} P.~M.,  2008, \mn@doi [\mnras]
  {10.1111/j.1365-2966.2007.12631.x}, \href
  {http://adsabs.harvard.edu/abs/2008MNRAS.383.1195W} {383, 1195}

\bibitem[\protect\citeauthoryear{Zarka et~al.,}{Zarka
  et~al.}{2015}]{zarka:hal-01196457}
Zarka P.,  et~al., 2015, in {2015 International Conference on Antenna Theory
  and Techniques (ICATT)}. 2015 International Conference on Antenna Theory and
  Techniques (ICATT).
Kharkiv, Ukraine, \mn@doi{10.1109/ICATT.2015.7136773}, \url
  {https://hal.archives-ouvertes.fr/hal-01196457}

\bibitem[\protect\citeauthoryear{{Zhang}, {Zuo}, {Ansari}, {Chen}, {Li}, {Wu},
  {Campagne}  \& {Magneville}}{{Zhang} et~al.}{2016a}]{2016RAA....16..158Z}
{Zhang} J.,  {Zuo} S.-F.,  {Ansari} R.,  {Chen} X.,  {Li} Y.-C.,  {Wu} F.-Q.,
  {Campagne} J.-E.,   {Magneville} C.,  2016a, \mn@doi [Research in Astronomy
  and Astrophysics] {10.1088/1674-4527/16/10/158}, \href
  {https://ui.adsabs.harvard.edu/abs/2016RAA....16..158Z} {16, 158}

\bibitem[\protect\citeauthoryear{{Zhang}, {Ansari}, {Chen}, {Campagne},
  {Magneville}  \& {Wu}}{{Zhang} et~al.}{2016b}]{2016MNRAS.461.1950Z}
{Zhang} J.,  {Ansari} R.,  {Chen} X.,  {Campagne} J.-E.,  {Magneville} C.,
  {Wu} F.,  2016b, \mn@doi [\mnras] {10.1093/mnras/stw1458}, \href
  {http://adsabs.harvard.edu/abs/2016MNRAS.461.1950Z} {461, 1950}

\bibitem[\protect\citeauthoryear{{van Haarlem} et~al.,}{{van Haarlem}
  et~al.}{2013}]{2013A&A...556A...2V}
{van Haarlem} M.~P.,  et~al., 2013, \mn@doi [\aap]
  {10.1051/0004-6361/201220873}, \href
  {http://esoads.eso.org/abs/2013A%26A...556A...2V} {556, A2}

\bibitem[\protect\citeauthoryear{{van Weeren} et~al.,}{{van Weeren}
  et~al.}{2016}]{2016ApJS..223....2V}
{van Weeren} R.~J.,  et~al., 2016, \mn@doi [\apjs] {10.3847/0067-0049/223/1/2},
  \href {https://ui.adsabs.harvard.edu/abs/2016ApJS..223....2V} {223, 2}

\makeatother
\end{thebibliography}

\bsp	
\label{lastpage}
\end{document}